# Broadband Single-Particle Absorption Circular Dichroism Reveals Chiroptical Heterogeneity in Gold Helicoids

**Rohit B. Raj,[1] Susanna Bertuletti,[2] Jeong Hyun Han,[3] Sepin Cho,[3] Debapriya Pal,[4] Nick Feldman,[4] Ki Tae Nam,[3] Willem L. Noorduin,[2,5] A. Femius Koenderink,[4,5] Erik C. Garnett[*1,5]**

**[1] LMPV-Sustainable Energy Materials Department, AMOLF Institute, Science Park 104, Amsterdam, 1098XG, The Netherlands**

**[2] Autonomous Matter, AMOLF, Science Park 104, Amsterdam, 1098XG, The Netherlands**

**[3] Department of Materials Science and Engineering, Seoul National University, Seoul 08826, Republic of Korea**

**[4] Department of Information in Matter and Center for Nanophotonics, AMOLF, Science Park 104, Amsterdam, 1098XG, The Netherlands**

**[5] University of Amsterdam, Science Park 904, Amsterdam, 1098XH, The Netherlands**

**[*]Corresponding author. Email: e.garnett@amolf.nl**

**Abstract**

Quantitative measurements of absorption circular dichroism (CD) at the single-particle level are essential for understanding how individual chiral nanostructures dissipate optical energy, yet broadband absorptance measurements remain experimentally challenging. Here, we introduce a wavelength-tunable integrating-sphere microscope that determines the absorptance of individual nanoparticles under right- and left-circularly polarized illumination through direct optical energy balance, enabling broadband measurement of the absorption dissymmetry factor, $g_{abs}$. Measurements of achiral gold nanospheres and strongly anisotropic gold nanorods establish the apparent absorption CD baseline and demonstrate minimal detectable linear-dichroism-to-circular-dichroism (LD-to-CD) leakage under the experimental conditions. Applying this approach to 87 chiral gold helicoids synthesized using L-glutathione (L-helicoids) and 96 synthesized using D-glutathione (D-helicoids) reveals mean particle-level $g_{abs}$ values of opposite sign, with a statistically significant difference between the two populations. Individual particles nevertheless exhibit pronounced heterogeneity in response sign, magnitude, spectral position, and line shape. More than one-third of the particles in each population display opposite-sign absorption CD responses relative to their population-average tendency. Correlative SEM analysis of the L-helicoid population further shows that opposite-sign responses persist among isolated particles exhibiting the characteristic projected helicoid morphology, indicating that aggregation and gross differences in projected morphology are insufficient to explain the observed heterogeneity. These results establish broadband single-particle absorption CD spectroscopy as a direct probe of absorptive chiroptical heterogeneity and reveal particle-specific responses obscured by ensemble averaging.

## Introduction

Chiral plasmonic nanoparticles can exhibit large and wavelength-dependent differences in their interactions with left-circularly polarized (LCP) and right-circularly polarized (RCP) light, enabling applications in polarization control,[1–3] chiral sensing,[4] enantioselective separation,[5,6] asymmetric catalysis,[7–9] and polarization-dependent photochemistry.[10,11] Their chiroptical responses can arise through both scattering and absorption, which govern distinct physical processes. Scattering redirects optical energy into the far field, whereas absorption describes optical energy dissipated within the nanoparticle. Unlike extinction, which combines absorption and scattering, absorption directly probes the energy available for photothermal heating, hot-carrier generation, and photochemical processes.[12–14] Because extinction contains both absorptive and scattering contributions, an extinction circular dichroism (CD) measurement does not uniquely determine the absorptive chiroptical response; the two channels can exhibit different dissymmetry magnitudes and spectral line shapes. Absorption CD, defined here as the differential absorptance under LCP and RCP illumination, is therefore a distinct and physically important quantity. This response is quantified by the absorption dissymmetry factor, $g_{abs}$.

The distinction between absorption and scattering g-factors becomes particularly important at the single-particle level. Advances in colloidal synthesis have produced three-dimensional chiral plasmonic nanoparticles with large ensemble dissymmetry factors and increasingly well-controlled chiral morphologies.[15–17] Beyond achieving controlled morphology, a central challenge is to determine whether particles belonging to the same chiral population exhibit consistent absorptive chiroptical responses. Chemically synthesized particles inevitably vary in size, shape, surface faceting, and crystallinity, and even subtle variations in these features can shift plasmonic resonances and alter their chiroptical line shapes.[18,19] Ensemble CD spectra report only the population average and conceal the underlying distribution of particle-specific responses. Recent broadband single-particle scattering CD measurements of gold helicoids have indeed revealed substantial heterogeneity, with some particles exhibiting scattering dissymmetry factors, $g_{scat}$, several times larger than the corresponding ensemble values.[20] Because radiative and nonradiative channels need not follow the same chiroptical trends, it remains unclear how reliably the chiral growth conditions translate into particle-level absorption CD and whether the projected morphology can account for deviations from the population-average tendency.

The first experimental challenge is obtaining quantitative, wavelength-resolved absorption CD spectra of individual nanoparticles across a broad spectral range. Existing broadband single-particle methods primarily probe extinction or scattering CD: extinction combines absorption and scattering, whereas dark-field measurements selectively detect the radiative response.[20–23] Photothermal CD microscopy provides a direct and highly sensitive, absorption-selective probe.[24] Because the signal is generated through optical heating, the excitation conditions must be chosen to balance photothermal sensitivity against possible heat-induced changes to the nanostructure.[25,26] Most reported single-particle photothermal CD measurements have focused on one or a few selected excitation wavelengths rather than on complete broadband spectra.[24–27] For populations with particle-dependent resonance positions and line shapes, predetermined wavelengths may not coincide with the maximum absorption dissymmetry of every particle. Quantitative broadband measurements are therefore needed to recover the full absorption CD spectrum and determine how absorptive chirality is distributed across individual nanoparticles. Despite advances in extinction, scattering, and photothermal chiroptical microscopy, quantitative broadband absorption CD spectra of individual nanoparticles have not yet been demonstrated.

A second experimental challenge is the rigorous exclusion of polarization artifacts.[26,28,29] Linear dichroism (LD), defined as differential absorption under orthogonal linear polarizations, can be substantially larger than CD for an individual anisotropic nanoparticle.[28,29] Real particles deviate from ideal rotational symmetry and may therefore exhibit nonzero LD even when their morphology is highly symmetric. If the LCP and RCP illumination states contain unequal residual linear-polarization components, part of this LD response can leak into their difference and appear as a CD signal. Recent single-particle circular differential scattering studies have shown that wave-plate imperfections and polarization distortions can couple particle anisotropy into spurious chiroptical responses in achiral nanostructures.[28,29] Although strategies for identifying and correcting such artifacts have been developed for scattering measurements,[28,29] equivalent validation has not yet been established for quantitative broadband single-particle absorptance spectroscopy.

Here, we establish quantitative broadband absorption CD spectroscopy of individual plasmonic nanoparticles using a wavelength-tunable integrating-sphere microscope. By measuring reflection and the combined transmitted-plus-scattered power, the method determines absorptance directly through optical energy balance.[30] Measurements of isotropic gold nanospheres and strongly anisotropic gold nanorods establish the instrumental apparent absorption CD baseline and demonstrate minimal detectable LD-to-CD leakage under our experimental conditions. We then apply the method to 87 L-helicoids and 96 D-helicoids, chiral gold Helicoid III nanoparticles synthesized using an established glutathione-directed growth method, with L-glutathione (L-GSH) and D-glutathione (D-GSH) used as the respective chiral growth-directing agents.[19] The two populations exhibit mean particle-level $g_{abs}$ values of opposite sign, yet their single-particle responses are broadly distributed and substantially overlapping in sign, magnitude, spectral position, and line shape. More than one-third of each population exhibits a response opposite in sign to its population-average tendency. A morphology-resolved analysis of the L-helicoid population further shows that these opposite-sign responses persist among isolated particles exhibiting the characteristic projected helicoid morphology. Thus, the chiral growth conditions used to synthesize the L- and D-helicoid populations govern their average absorptive chiroptical tendency but do not uniquely determine the chiroptical response of an individual gold helicoid. Moreover, projected SEM morphology alone is insufficient to account for the observed particle-to-particle heterogeneity.

## Results and Discussion

### Broadband Single-Particle Absorption Circular Dichroism

Absorption CD occurs when a chiral nanoparticle absorbs different fractions of the incident optical power under RCP and LCP illumination, such that $A_{RCP} \neq A_{LCP}$ (Figure 1a). We quantify this difference using the absorption dissymmetry factor

$$g_{abs}(\lambda) = \frac{2[A_{RCP}(\lambda) - A_{LCP}(\lambda)]}{A_{RCP}(\lambda) + A_{LCP}(\lambda)},$$

where positive and negative values indicate preferential absorption under RCP and LCP illumination, respectively.

To measure the polarization-dependent absorptance of individual nanoparticles, we used the wavelength-tunable integrating-sphere microscope illustrated schematically in Figure 1b. Wavelength-selected light was passed through a linear polarizer. After the beam splitter, a quarter-wave plate (QWP) was used to generate RCP or LCP polarization for absorption CD measurements, or a half-wave plate (HWP) to generate horizontal (H) or vertical (V) polarization for linear dichroism measurements. Along the incident-light path, the beam splitter directed a fraction of the optical power to the beam-monitor photodetector (omitted from the simplified schematic in Figure 1b for clarity), while the remaining light propagated toward the sample and was focused onto an individual nanoparticle using an NA 0.42 objective. Light reflected from the sample propagated back through the objective and was directed by the same beam splitter to the reflection photodetector. Transmitted and scattered light were collected by the integrating sphere. The beam-monitor signal was used to correct the detector signals for temporal fluctuations in the incident optical power. The complete optical layout and detector-normalization procedure are provided in Section 1 of the SI.

The generated polarization states were independently characterized at the sample plane using a commercial polarimeter. The circular polarization states exhibited $|S_3| > 0.986$ throughout the investigated wavelength range, indicating a high degree of circular polarization (see Section 1.1 and Figures S1b and S1c of the SI for the complete wavelength-dependent Stokes parameters).

For each incident polarization state p, the reflection photodetector recorded the reflected power, $P_R$, while the integrating-sphere photodetector recorded the combined transmitted and scattered power, $P_T + P_S$. The detector signals were calibrated and converted into the reflected fraction, $R_p$, and the combined transmitted-plus-scattered fraction, $(T + S)_p$, of the incident optical power (see Section 1.3 of the SI for details of the signal calibration). The absorptance under each polarization state p was then determined through optical energy balance:

$$A_p(\lambda) = 1 - R_p(\lambda) - (T + S)_p(\lambda), \quad p \in \{H, V, RCP, LCP\}.$$

This procedure yields wavelength-resolved $A_{RCP}$, $A_{LCP}$, and $g_{abs}$ spectra for individual nanoparticles.

A central challenge in single-particle CD measurements is the leakage of LD into the measured CD signal, which can produce an apparent CD response in achiral structures.[28,29] We therefore first benchmarked the measurement using isotropic achiral gold nanospheres and strongly anisotropic achiral gold nanorods. The nanospheres establish the apparent $g_{abs}$ baseline, whereas the nanorods provide a stringent test of LD-to-CD leakage.[28]

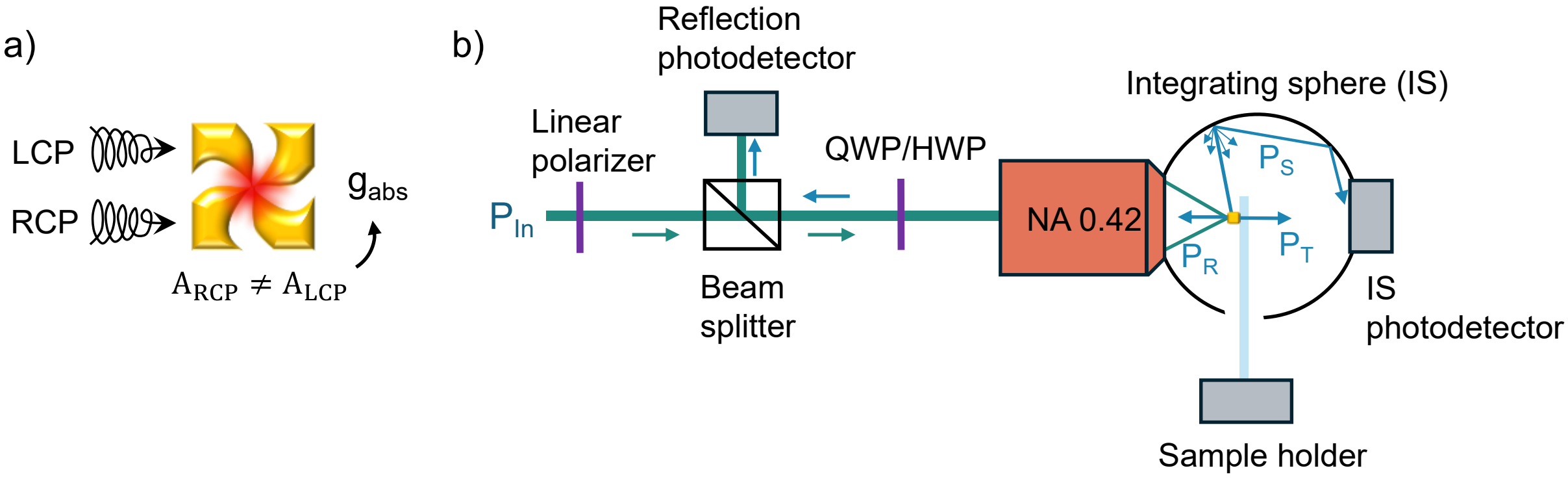

**Figure 1. Broadband single-particle absorption circular dichroism measurement.** (a) Schematic of a chiral nanoparticle exhibiting different absorptances under RCP and LCP illumination. (b) Simplified schematic of the wavelength-tunable integrating-sphere microscope. Wavelength-selected incident light of power $P_{In}$ passes through a linear polarizer and either a QWP or HWP before being focused onto an individual nanoparticle using an NA 0.42 objective. The reflection photodetector records the reflected power, $P_R$, while the integrating sphere collects the transmitted and scattered powers, $P_T$ and $P_S$, as a combined signal detected by the IS photodetector. A beam-monitor photodetector used to correct source-intensity fluctuations is omitted from the simplified schematic for clarity. Absorptance is determined from the calibrated reflected and combined transmitted-plus-scattered fractions through optical energy balance.

**Validation of Single-Particle Absorption CD and Assessment of LD-to-CD Leakage**

To establish the apparent CD baseline of the microscope and assess LD-to-CD leakage, we measured 16 isotropic achiral gold nanospheres with a diameter of 100 nm and 11 strongly anisotropic achiral gold nanorods (40 nm × 80 nm). The nanospheres provide an isotropic achiral reference for the apparent CD baseline, whereas the nanorods are also achiral but exhibit strong linear dichroism because of their shape anisotropy. They therefore provide a stringent control for determining whether LD leaks into the measured CD signal. We drop-cast dilute nanoparticle dispersions onto glass substrates to obtain spatially isolated particles with interparticle separations exceeding 3 μm, as shown in the representative dark-field images in Figures S2c and S3c of the SI (see Section 2 in the SI for sample-preparation details). Individual nanoparticles were first measured optically and subsequently correlated with their scanning electron microscopy (SEM) images.

A representative gold nanosphere is shown in Figure 2a. Its absorptance spectrum exhibits a localized surface plasmon resonance (LSPR) near 550 nm, with the $A_{RCP}$ and $A_{LCP}$ spectra closely overlapping across the resonance. This yields $g_{abs} = 0.00 \pm 0.03$ for the representative nanosphere, where the uncertainty represents the standard deviation within the selected spectral region (see Section 4.1 of the SI for the $g_{abs}$ calculation and peak-selection procedure). This response establishes the apparent CD baseline measured for an isotropic achiral nanoparticle.

Gold nanorods provide a more stringent test of LD-to-CD leakage because their absorptance depends strongly on the orientation of the incident electric field relative to the rod long axis.[31–33] The longitudinal LSPR couples most strongly to an electric field aligned with the rod long axis and therefore produces a pronounced difference between absorption under H and V polarization.[34] We quantify this linear anisotropy using the linear-dissymmetry factor

$$g_{LD}(\lambda) = \frac{2\,[A_H(\lambda) - A_V(\lambda)]}{A_H(\lambda) + A_V(\lambda)},$$

where positive and negative values indicate preferential absorption under H and V polarization, respectively. Representative nanorods oriented predominantly along the H and V axes are shown in Figures 2b and 2c. For the horizontally oriented rod in Figure 2b, $E_H$ is approximately aligned with the rod long axis and therefore couples strongly to the longitudinal LSPR, resulting in substantially larger $A_H$ than $A_V$ and a positive $g_{LD}$ (see Section 4.1 of the SI for the $g_{LD}$ calculation and peak-selection procedure). Conversely, for the vertically oriented rod in Figure 2c, $E_V$ is approximately aligned with the rod long axis, giving a stronger $A_V$ response and a negative $g_{LD}$. A nanorod with an intermediate orientation, for which both $A_H$ and $A_V$ contribute substantially, is shown in Figure S5 of the SI. Despite these pronounced linear-polarization responses, the $A_{RCP}$ and $A_{LCP}$ spectra remain closely overlapping for the nanorods, resulting in $g_{abs}$ values near zero.

The population-level responses are summarized in Figures 2d and 2e for the 16 nanospheres and 11 nanorods, respectively. The nanosphere resonances cluster near 550 nm, whereas the longitudinal resonances of the nanorods span a broader, red-shifted wavelength range. This broader distribution in nanorods is expected because the longitudinal LSPR of a nanorod is highly sensitive to its aspect ratio,[34] such that modest particle-to-particle variations in rod dimensions can produce appreciable shifts in the resonance wavelength. The population-averaged $g_{abs}$ values are $-0.02 \pm 0.05$ for the nanospheres and $0.02 \pm 0.05$ for the nanorods, with no systematic dependence on resonance wavelength for either control population. Figure 2f directly compares the circular- and linear-dissymmetry of individual nanorods. For each rod, its $g_{abs}$ and $g_{LD}$ values are connected with a dashed line. The rods exhibit strong linear anisotropy, with a mean $|g_{LD}|$ of $1.08 \pm 0.52$, whereas the mean $|g_{abs}|$ is only $0.05 \pm 0.02$. Thus, despite individual nanorods exhibiting large positive or negative $g_{LD}$ values depending on their orientation, their $g_{abs}$ values remain clustered near zero. Together, these achiral controls define the experimentally observed apparent $g_{abs}$ baseline and demonstrate minimal detectable LD-to-CD leakage under our measurement conditions.

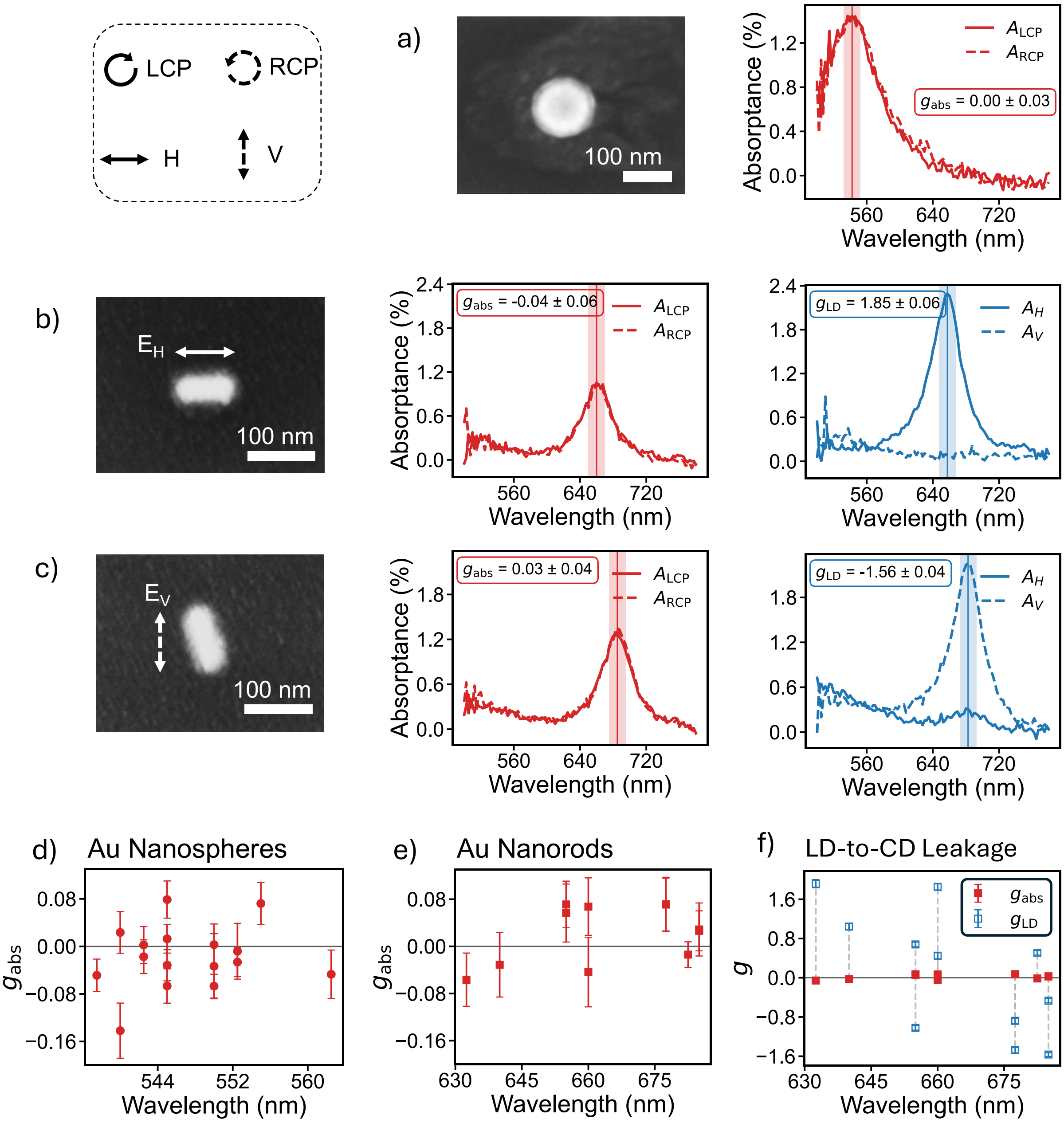

**Figure 2. Validation of absorption CD with achiral Au nanospheres and nanorods**. (a) Isotropic 100 nm Au nanosphere control, showing the SEM image and $A_{RCP}$ and $A_{LCP}$ spectra with the corresponding $g_{abs}$ value. Anisotropic Au nanorod controls (40 nm × 80 nm) oriented predominantly along the (b) H and (c) V polarization axes, showing the corresponding SEM images, circular-polarization absorptance spectra and $g_{abs}$ values, and linear-polarization absorptance spectra and $g_{LD}$ values. The $E_H$ and $E_V$ arrows indicate the incident electric-field directions. Shaded regions in (a)–(c) indicate the 20 nm-wide spectral windows used to determine the reported $g_{abs}$ and $g_{LD}$ values. Population-level $g_{abs}$ as a function of resonance wavelength for

(d) 16 Au nanospheres and (e) 11 Au nanorods. (f) Direct comparison of $g_{abs}$ and $g_{LD}$ for individual nanorods to assess LD-to-CD leakage.

## From Ensemble Chirality to Single-Particle Absorption CD

Having established negligible detectable LD-to-CD leakage using the achiral controls, we next investigated L- and D-helicoids. The particles have characteristic lateral dimensions of approximately 180 nm and consist of four twisted arms arranged with approximate fourfold rotational symmetry about each face. Concave gaps separate the arms, with the two populations differing in the direction of the arm twist (see Section 5 of the SI for synthesis details). The helicoid dispersions were drop-cast onto glass substrates at sufficiently low surface density to obtain spatially isolated particles for single-particle measurements (see Section 5.3 of the SI for deposition details and representative dark-field images).

The ensemble extinction spectra of the L- and D-helicoid colloidal dispersions are shown in Figure 3a, with the corresponding extinction dissymmetry spectra, $g_{ext}$, shown in Figure 3b. The $g_{ext}$ spectra were measured using a commercial CD spectrometer (see Section 5.2 of the SI for measurement details). Both dispersions exhibit broad plasmonic extinction extending across the visible and near-infrared regions. Their strongest ensemble chiroptical responses occur around 590–630 nm, where the L-helicoid population exhibits a negative $g_{ext}$ of approximately −0.18, while the D-helicoid population exhibits a positive $g_{ext}$ of approximately +0.13 (Figure 3b). These opposite signs establish the expected population-average chiroptical tendency of the L- and D-helicoids.

The chiroptical response of related helicoids has been associated with hybrid multipolar plasmon modes and strong electromagnetic coupling within the concave chiral gaps.[35,36] We therefore qualitatively associate the ensemble response in this spectral region with similar gap-coupled chiral plasmon modes, while recognizing that the exact resonance positions and modal contributions depend on particle geometry and dielectric environment. Because extinction contains both absorption and scattering contributions, the ensemble spectra $g_{ext}$ provide a reference for the expected response sign but cannot be compared quantitatively on a one-to-one basis with $g_{abs}$.

To determine whether the opposite-sign chiroptical responses observed in the ensemble measurements are also present in the absorptive response of individual particles, we measured individual L- and D-helicoids optically and then imaged the same particles by SEM to correlate their optical responses with projected morphology. Figures 3c and 3d show representative particles with opposite directions of arm twist in their projected morphologies. Their polarization-dependent absorptance spectra are shown in Figures 3e and 3f. Both particles exhibit two prominent absorptance features across the measured spectral range. The red-shifted feature shows the clearest difference between $A_{LCP}$ and $A_{RCP}$, with the direction of this difference reversing between the representative L- and D-helicoids (Figures 3e and 3f). Correspondingly, their $g_{abs}$ spectra exhibit pronounced responses of opposite sign in this spectral region (Figures 3g and 3h).

For the representative L-helicoid, $A_{LCP}$ exceeds $A_{RCP}$ within the selected spectral region (Figure 3e), producing a negative response of $g_{abs} = -0.46 \pm 0.03$ (Figure 3g). Conversely, the representative D-helicoid exhibits stronger absorption under RCP illumination (Figure 3f), yielding $g_{abs} = +0.29 \pm 0.03$ (Figure 3h). For both particles, the uncertainty represents the standard deviation within the selected shaded spectral region (see Section 6 of the SI for the $g_{abs}$ calculation). The opposite signs of these representative single-particle responses are consistent with the ensemble-average tendencies of the L- and D-helicoids.

These measurements demonstrate that broadband single-particle measurements can resolve strong wavelength-dependent absorption CD responses from individual gold helicoids. However, chemically synthesized helicoids can vary in size, arm development, surface curvature, and chiral-gap geometry, potentially producing substantial variation in their absorptive chiroptical response. Previous single-particle scattering and photothermal CD studies of related helicoids and other chiral plasmonic nanoparticles have likewise reported pronounced particle-to-particle variability.[20,25] Measurements of one representative particle from each synthesis population therefore cannot establish how consistently the ensemble-average tendency is retained across the population or how broadly $g_{abs}$ is distributed. To address these questions, we extended the measurements to 87 L-helicoids and 96 D-helicoids.

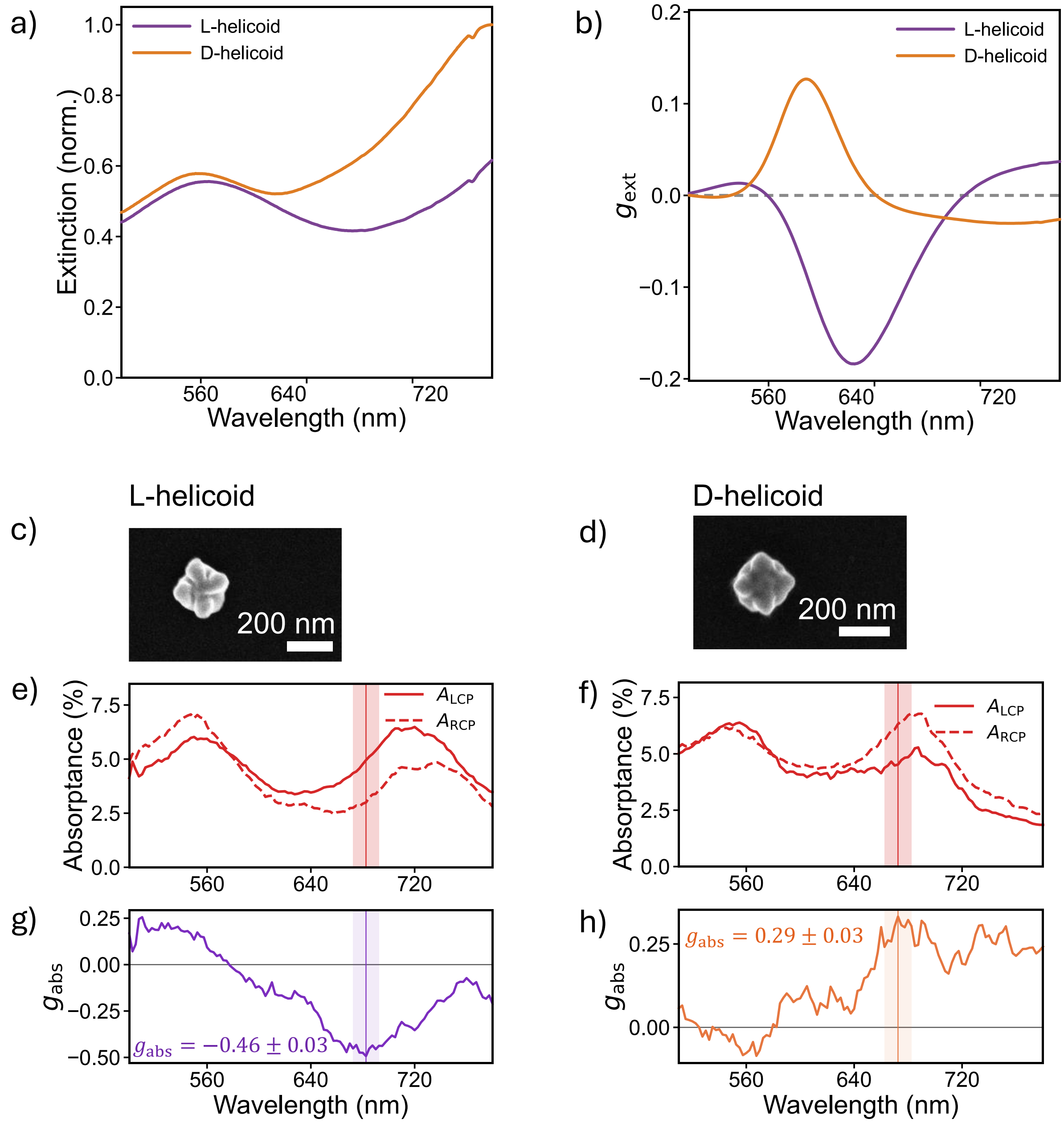

**Figure 3. Ensemble and representative single-particle chiroptical responses of gold helicoids.** Ensemble (a) extinction spectra and (b) corresponding extinction dissymmetry

spectra, $g_{ext}$, of L- and D-helicoid dispersions. SEM images of representative (c) L-helicoid and (d) D-helicoid. Polarization-dependent absorptance spectra of the representative (e) L-helicoid and (f) D-helicoid, with the corresponding $g_{abs}$ spectra shown in (g) and (h), respectively. Shaded regions in (e)–(h) indicate the 20 nm-wide spectral windows used to determine the reported $g_{abs}$ values.

**Population-Level Heterogeneity in Single-Particle Absorption Circular Dichroism**

To determine whether the representative responses in Figure 3 reflect the broader L- and D-helicoid populations, we measured wavelength-resolved absorption CD spectra from 87 L-helicoids and 96 D-helicoids. Figures 4a and 4b show the overlaid $g_{abs}$ spectra of the L- and D-helicoids, respectively, with each trace representing one particle. Both populations exhibit substantial particle-to-particle variation in the magnitude, spectral position, and detailed line shape of the absorption CD response. Individual spectra can also change sign across different spectral regions, indicating that the absorptive chiroptical response of a given particle is wavelength-dependent.

The corresponding population-averaged $g_{abs}$ spectra are shown in Figures 4c and 4d. Around 660 nm, the L-helicoid population exhibits a predominantly negative average response, whereas the D-helicoid population exhibits a predominantly positive average response. These opposite signs are consistent with the corresponding dominant ensemble $g_{ext}$ tendencies observed in Figure 3b, establishing opposite average absorptive chiroptical tendencies for the two populations around this red-shifted feature. At shorter wavelengths, however, both population averages exhibit a positive response. The ensemble $g_{\text{ext}}$ spectra likewise depart from the ideal enantiomeric relationship in this wavelength region (Figure 3b). For ideal geometrically mirror-image helicoids in identical environments, the chiroptical spectra would be expected to exhibit opposite signs at a given wavelength;[37] the common-sign response therefore represents a departure from this ideal mirror relationship and may reflect structural variability between the chemically synthesized populations, which can shift the spectral positions and relative strengths of the plasmonic modes contributing to the chiroptical response.[37–39]

Despite the clear opposite-sign population tendency of $g_{abs}$ in the red-shifted spectral region, the population-averaged spectra mask substantial heterogeneity at the single-particle level. Particle-to-particle variations in spectral position and line shape broaden features during wavelength-by-wavelength averaging, while responses of opposite sign partially cancel. Consequently, strong absorption CD responses from individual particles can be substantially reduced in the population-averaged spectra. To examine this heterogeneity at the single-particle level, we analyzed the particle-level $g_{abs}$ distributions for the L- and D-helicoid populations (see Section 6 of the SI for the analysis procedure). The resulting distributions are shown in Figures 4e and 4f, respectively. Strikingly, both populations contain substantial numbers of particles with positive and negative $g_{abs}$ responses. The gray region indicates the achiral-control reference range of $\pm 0.06$, established from the pooled nanosphere and nanorod controls. Relative to this reference range, particle responses were classified as expected-sign, within the reference range, or opposite-sign. Expected-sign responses were those exceeding the reference range in the direction of the corresponding population-average tendency, whereas opposite-sign responses exceeded the range in the opposite direction.

Within the L-helicoid population, 58.6% of the particles exhibit an expected-sign response, 4.6% fall within the achiral-control reference range, and 36.8% exhibit an opposite-sign response. Within the D-helicoid population, 63.5% exhibit an expected-sign response, none fall within the achiral-control reference range, and 36.5% exhibit an opposite-sign response. Accordingly, more than one-third of the particles in each population exhibit an absorption CD response opposite in sign to the corresponding population-average tendency. Notably, very few particles fall within the achiral-control range; instead, most exhibit a pronounced chiroptical response of either sign. The rarity of near-achiral responses shows that the measured populations are dominated by chiroptically active particles. This observation is consistent with the sensitivity of structurally complex plasmonic nanoparticles to small particle-specific asymmetries, which can give rise to appreciable chiroptical responses at the single-particle level. Such sensitivity to small structural deviations has also been reported for complex plasmonic nanoparticles designed to be nominally achiral.[27,28]

Despite this substantial heterogeneity, the two distributions remain shifted in opposite directions. The L- and D-helicoid populations have mean $g_{abs}$ values of −0.11 and +0.10, respectively, and the difference between the population means is statistically significant according to a two-sided Welch independent-samples t-test ($p = 7.99 \times 10^{-5}$). The L- and D-directed syntheses therefore bias the average absorptive chiroptical responses in opposite directions but do not uniquely determine the sign or magnitude of $g_{abs}$ for an individual helicoid. The sign of $g_{abs}$ reflects which circular polarization is preferentially absorbed by a particle and therefore depends on its particle-specific electromagnetic response at a given wavelength. Particle-to-particle variations in three-dimensional structures can modify the spectral position, strength, and coupling of the underlying plasmonic modes,[20,25,27] providing a plausible origin of the observed variability. We therefore tested whether readily observable structural differences, including aggregation or conspicuous differences in projected particle morphology, could account for the opposite-sign responses.

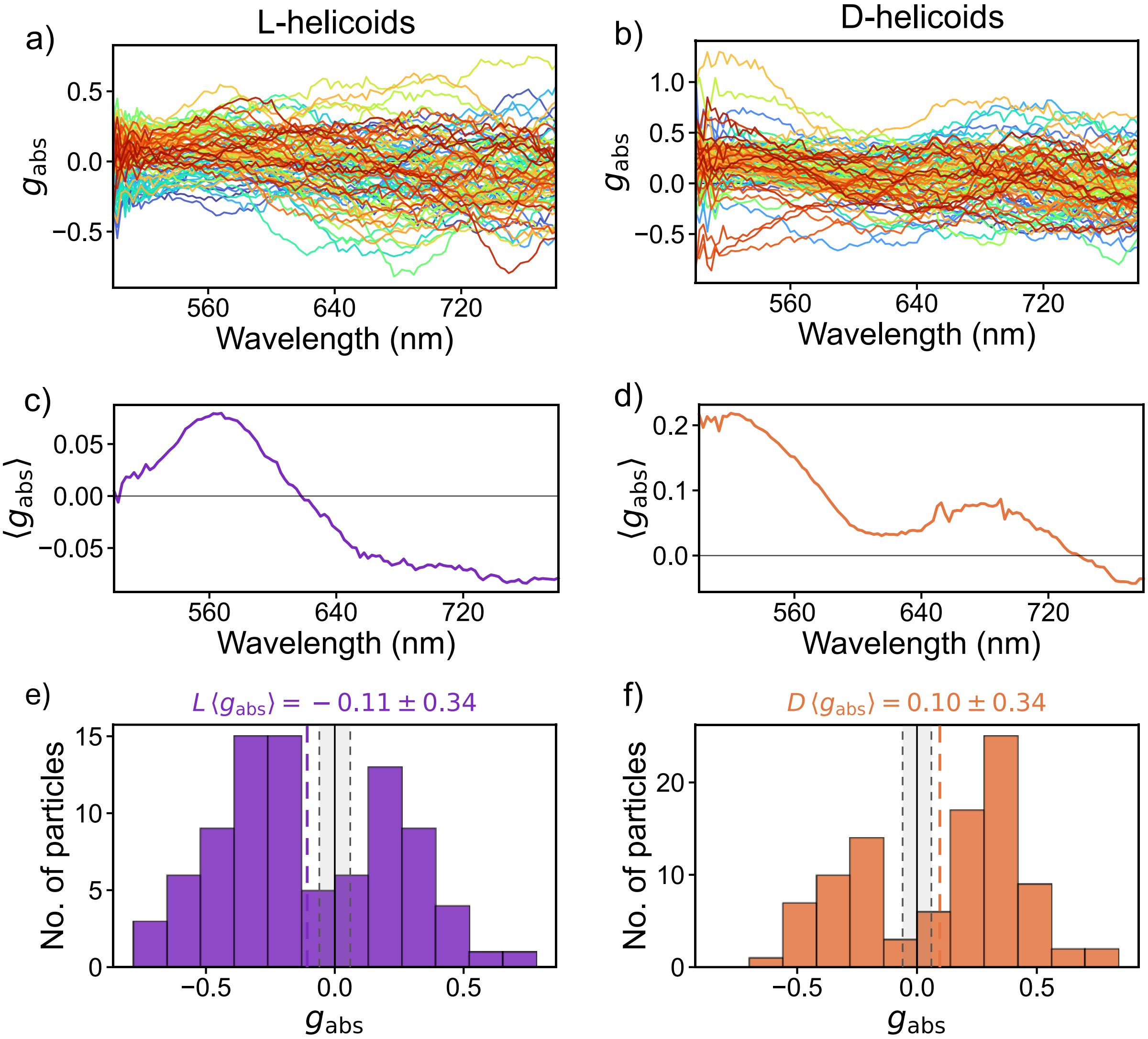

**Figure 4. Population-level heterogeneity in single-particle absorption circular dichroism of gold helicoids.** Wavelength-resolved $g_{abs}(\lambda)$ spectra of all measured (a) L-helicoids ($N = 87$) and (b) D-helicoids ($N = 96$), with each trace representing one particle. Corresponding population-averaged $g_{abs}(\lambda)$ spectra for the (c) L-helicoid and (d) D-helicoid populations. Horizontal gray

lines indicate $g_{abs} = 0$. Distributions of $g_{abs}$ values for the (e) L-helicoids and (f) D-helicoids. The gray shaded region in (e) and (f) indicates the achiral-control reference range of $g_{abs} = \pm 0.06$, established from the pooled nanosphere and nanorod controls.

## Projected Morphology Does Not Uniquely Account for Single-Particle Absorption CD

The population analysis in Figure 4 revealed substantial heterogeneity in both the sign and magnitude of the absorption CD response within the L- and D-helicoid populations. To assess whether this variability could be attributed to aggregation or conspicuous differences in particle morphology, we performed a morphology-resolved analysis of the L-helicoid population (see Section 7 of the SI for details); representative SEM images of D-helicoids are provided in Figure S7 in the SI for comparison. Accordingly, all morphology-resolved data in Figure 5 refer to L-helicoids.

The SEM-correlated particles were classified according to their projected morphology as Characteristic singles, Atypical singles, and Aggregates. Characteristic singles displayed the projected helicoid morphology of the L-helicoids, with well-developed twisted arms, clearly defined concave gaps, and approximate fourfold rotational symmetry. Atypical singles did not display this characteristic helicoid shape in their SEM projection. Aggregates comprised two or more particles in direct contact, including dimers and higher-order assemblies. The classification was performed from the SEM images without reference to the corresponding optical response (see Section 7 of the SI for details of the SEM correlation and morphology-classification procedure). Figure 5a shows the particle-level $g_{abs}$ values for the three morphology categories, with the gray region indicating the previously defined achiral-control reference range of $\pm 0.06$ in $g_{abs}$.

All three morphology categories span a broad range of $g_{abs}$ values, with both expected- and opposite-sign responses occurring in each category. Opposite-sign responses are observed in 35.9% of the Characteristic singles, despite these particles retaining the projected morphology expected for the L-helicoids. Opposite-sign responses therefore persist after excluding both Atypical singles and Aggregates. The opposite-sign fractions were similarly substantial for Atypical singles (39.3%) and lower for Aggregates (28.6%). Although the Aggregates were descriptively shifted further in the expected-sign direction, no statistically significant difference in $g_{abs}$ was detected among Characteristic singles, Atypical singles, and Aggregates using a Kruskal-Wallis test ($p = 0.549$). Consistent with this result, no statistically significant association was observed between morphology category and the occurrence of an opposite-sign response (exact test, $p = 0.825$) (see Section 7 of the SI for details of the statistical analyses).

Representative SEM-correlated particles illustrate these observations in Figures 5b–e. The Characteristic single shown in Figure 5b exhibits an expected-sign response, whereas a second Characteristic single with a broadly similar projected morphology exhibits an opposite-sign response (Figure 5c). The Atypical single in Figure 5d and the Aggregate in Figure 5e both exhibit expected-sign responses. Together with the population-level data in Figure 5a, these examples show that opposite-sign responses are not confined to Atypical singles or Aggregates but also occur among particles retaining the characteristic projected helicoid morphology.

Taken together, the morphology-resolved population analysis and representative particle measurements reveal no simple one-to-one correspondence between projected particle morphology and single-particle absorption CD. Most importantly, the persistence of opposite-sign responses among Characteristic singles demonstrates that aggregation and conspicuous morphological variations visible in SEM are insufficient to account for the observed heterogeneity. This does not exclude a structural origin because SEM provides only a two-dimensional projection of an intrinsically three-dimensional helicoid structure.

Particle-to-particle variations in the three-dimensional morphology therefore provide a plausible origin of the observed sign heterogeneity,[40–43] even among particles with similar projected morphologies.[20] More specifically, previous work on 432 helicoids has shown that their fourfold- and threefold-symmetry-related features can exhibit opposite geometric handedness, and that the relative prominence of these features is associated with the sign of the chiroptical response.[44] Differences in the relative development of these features could therefore alter the balance between chiral contributions to the optical response. The substrate-supported measurement geometry may provide an additional contribution by fixing the orientation of each particle relative to the incident beam,[20,44–46] thereby influencing which symmetry direction is preferentially probed. Differences in particle–substrate contact and the asymmetric dielectric environment at the glass interface may further perturb the plasmonic response.[47,48]

Correlative three-dimensional electron tomography combined with particle-specific electromagnetic modelling could therefore help identify the structural parameters governing the sign and magnitude of $g_{abs}$.[27,40,41,49] Establishing these structure-response relationships could, in turn, guide the synthesis of helicoid populations with more uniform and predictable absorptive chiroptical properties.

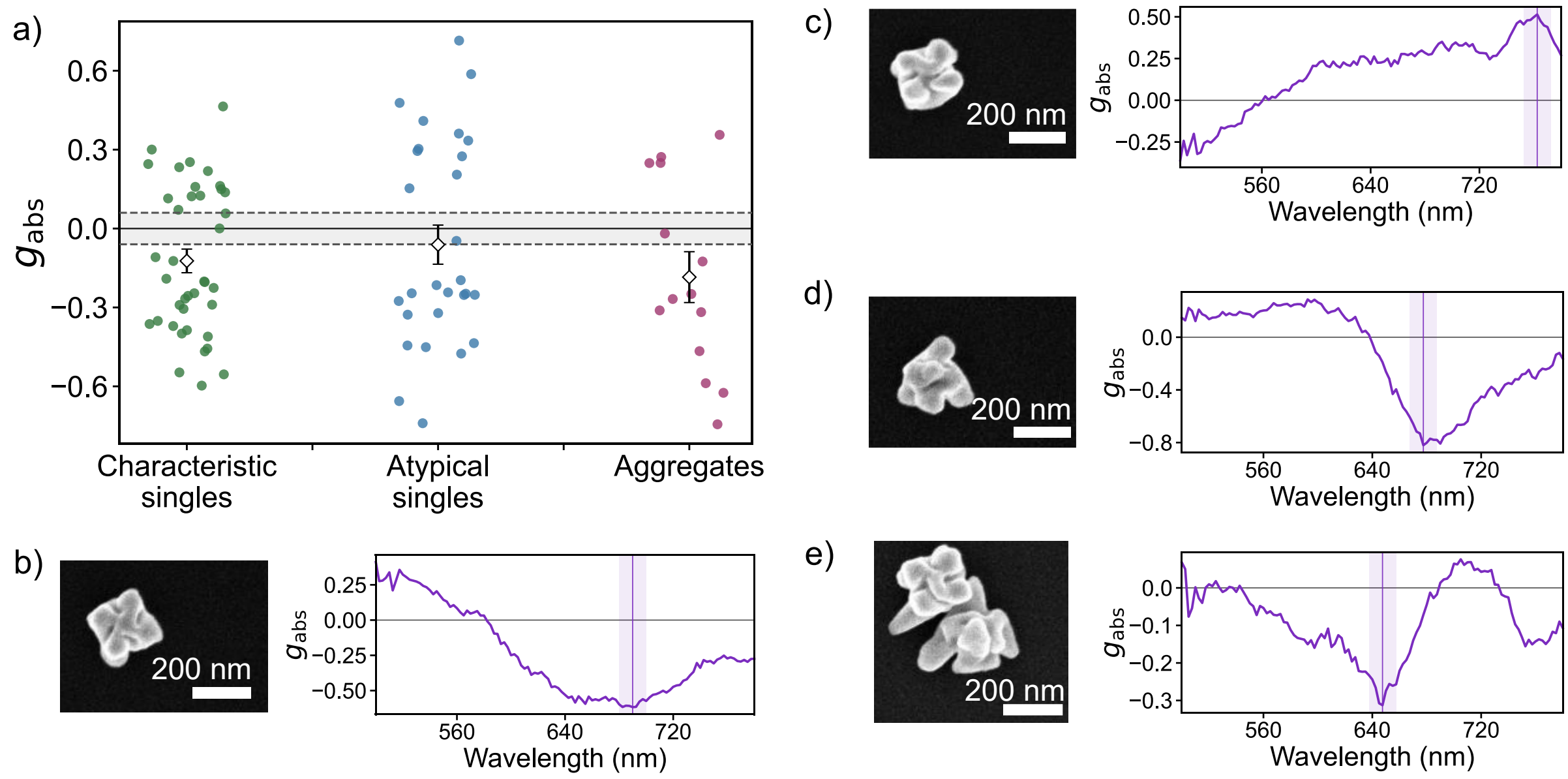

**Figure 5. Relationship between projected morphology and single-particle absorption circular dichroism in L-helicoids.** (a) Particle-level $g_{abs}$ values of SEM-correlated L-helicoids classified as Characteristic singles (N = 39), Atypical singles (N = 28), or Aggregates (N = 14). Each point represents one particle; points are horizontally jittered within each morphology category for visibility. The gray region denotes the achiral-control reference range, $-0.06 \leq g_{abs} \leq +0.06$, established from pooled nanosphere and nanorod controls. SEM images and corresponding $g_{abs}(\lambda)$ spectra of representative particles: (b) Characteristic single with the expected negative response, (c) Characteristic single with the opposite positive response, (d) Atypical single with the expected negative response, and (e) Aggregate with a negative response. Shaded regions in (b)–(e) indicate the 20 nm-wide spectral window used to determine the reported $g_{abs}$ values.

## Conclusions

We developed a wavelength-tunable integrating-sphere microscope that enables quantitative broadband absorption CD spectroscopy of individual nanoparticles through direct optical energy balance. Measurements of achiral gold nanospheres and strongly anisotropic gold nanorods established the apparent absorption CD baseline and demonstrated minimal detectable LD-to-CD leakage under the experimental conditions.

Applying this method to L- and D-helicoids revealed opposite population-average absorptive chiroptical tendencies and a statistically significant difference in mean particle-level $g_{abs}$ between the two populations. Individual particles nevertheless exhibited substantial heterogeneity in response sign, magnitude, spectral position, and line shape. More than one-third of the particles in each population displayed opposite-sign responses outside the achiral-control reference range. Thus, the chiral growth conditions used to synthesize the L- and D-helicoid populations govern their average absorptive chiroptical tendency but do not uniquely determine the $g_{abs}$ response of an individual helicoid.

Morphology-resolved analysis of the L-helicoids further showed that aggregation and conspicuous differences in projected SEM morphology are insufficient to account for this heterogeneity. Opposite-sign responses persisted among Characteristic singles displaying the expected projected helicoid morphology, demonstrating that gross morphological differences visible in a two-dimensional SEM projection do not uniquely account for the observed particle-to-particle variability.

These results establish broadband single-particle absorption CD spectroscopy as a direct approach for resolving absorptive chiroptical heterogeneity that is obscured by population averaging. Combining this method with correlative three-dimensional structural characterization and particle-specific electromagnetic modelling could help identify the nanoscale features governing absorptive chirality. Establishing such structure-response relationships could ultimately guide the design of chiral nanostructures for absorption-driven applications, including photothermal energy conversion, hot-carrier generation, photocatalysis, and polarization-dependent photochemistry.

**Supporting Information.** Wavelength-tunable integrating-sphere microscope and polarization characterization, particle localization and polarization-resolved spectral acquisition, detector normalization and absorptance calculation, preparation and characterization of achiral gold nanosphere and nanorod controls, additional nanorod LD-to-CD leakage analysis, extraction of $g_{\mathrm{abs}}$ and $g_{\mathrm{LD}}$ and determination of the achiral-control threshold, gold helicoid synthesis and sample preparation, particle-level $g_{\mathrm{abs}}$ extraction and population analysis, SEM-correlated morphology classification and statistical testing, and complete SEM catalogue of morphology-classified L-helicoids.

### Acknowledgments

We acknowledge financial support from the European Commission under the EIC Pathfinder CHIRALFORCE project (Grant Agreement No. 101046961). This work is also part of the Advanced Research Center for Chemical Building Blocks, ARC CBBC, which is co-founded and co-financed by the Dutch Research Council (NWO) and the Netherlands Ministry of Economic Affairs and Climate Policy. We thank Dr. Zihao Lu for contributions during the early stages of the project and for helpful discussions on morphology–chiroptical property correlations.

## For Table of Contents Use Only

# Broadband Single-Particle Absorption Circular Dichroism Reveals Chiroptical Heterogeneity in Gold Helicoids

**Rohit B. Raj,[1] Susanna Bertuletti,[2] Jeong Hyun Han,[3] Sepin Cho,[3] Debapriya Pal,[4] Nick Feldman,[4] Ki Tae Nam,[3] Willem L. Noorduin,[2,5] A. Femius Koenderink,[4,5] Erik C. Garnett*[1,5]**

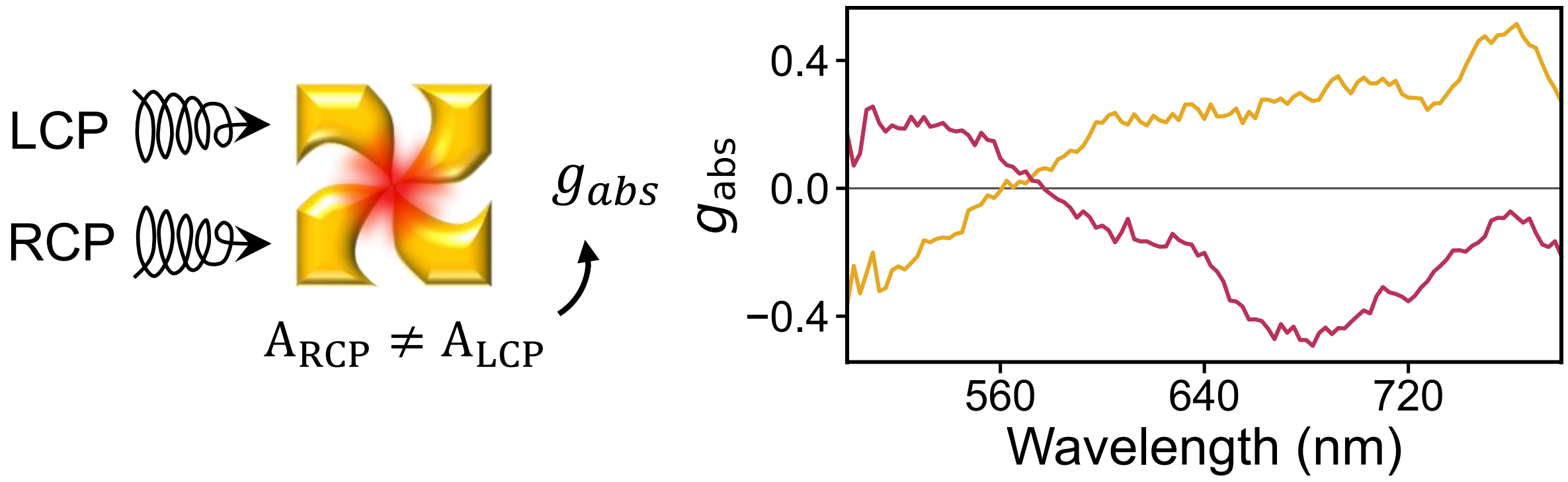

Broadband single-particle absorption circular dichroism reveals how individual chiral gold helicoids absorb LCP and RCP light differently. The resulting $g_{\mathrm{abs}}$ spectra show pronounced particle-to-particle variation, highlighting substantial chiroptical heterogeneity within the nanoparticle population.

# Supporting Information

## Broadband Single-Particle Absorption Circular Dichroism Reveals Chiroptical Heterogeneity in Gold Helicoids

**Rohit B. Raj,[1] Susanna Bertuletti,[2] Jeong Hyun Han,[3] Sepin Cho,[3] Debapriya Pal,[4] Nick Feldman,[4] Ki Tae Nam,[3] Willem L. Noorduin,[2,5] A. Femius Koenderink,[4,5] Erik C. Garnett[*1,5]**

**[1] LMPV-Sustainable Energy Materials Department, AMOLF Institute, Science Park 104, Amsterdam, 1098XG, The Netherlands**

**[2] Autonomous Matter, AMOLF, Science Park 104, Amsterdam, 1098XG, The Netherlands**

**[3] Department of Materials Science and Engineering, Seoul National University, Seoul 08826, Republic of Korea**

**[4] Department of Information in Matter and Center for Nanophotonics, AMOLF, Science Park 104, Amsterdam, 1098XG, The Netherlands**

**[5] University of Amsterdam, Science Park 904, Amsterdam, 1098XH, The Netherlands**

**[*]Corresponding author. Email: e.garnett@amolf.nl**

**Contents: 19 pages including cover page, 7 figures.**

## 1. Wavelength-Tunable Integrating-Sphere Microscope and Polarization Characterization

The wavelength-tunable integrating-sphere microscope used for the single-particle absorptance measurements is shown schematically in Figure S1a. The instrument is based on the integrating-sphere microscope described previously.[1] A supercontinuum laser (NKT Fianium FIU 15) was spectrally selected using an acousto-optic tunable filter (AOTF) operated with a Gooch & Housego AODS20200-8 driver. The wavelength-selected beam was passed through a mounted wire-grid polarizer (WP25M-VIS, Thorlabs) to define the initial linear-polarization state. A zero-order achromatic half-wave plate (AHWP10M-600, Thorlabs) or zero-order achromatic quarter-wave plate (AQWP05M-600, Thorlabs), positioned after the beam splitter, was used to generate H/V or RCP/LCP illumination, respectively.

A beam splitter directed a fraction of the incident beam toward a beam-monitor photodetector (Thorlabs PDA100A Si), while also separating light returning from the sample toward a dedicated reflection photodetector (Thorlabs PDA100A Si). The beam-monitor signal was used to normalize all simultaneously recorded detector signals, thereby correcting for temporal variations in the laser output and wavelength-dependent variations in the AOTF transmission.

The incident beam was focused onto the sample using an ultra-long-working-distance microscope objective (Mitutoyo M Apo Plan NIR 50×, NA 0.42, 17 mm working distance). The sample was positioned inside a modified GPS-020-SL integrating sphere (Labsphere) and mounted on a three-axis piezoelectric stage (Piezojena Tritor400). Light reflected or backscattered into the collection cone of the objective was directed by the beam splitter toward the reflection photodetector. Light transmitted through the substrate, together with light scattered into the integrating sphere, was redistributed by the Lambertian inner surface of the sphere and detected using an integrating-sphere photodetector (Newport 818-UV). The reflection detector therefore measured $P_R$, whereas the integrating-sphere detector measured the combined transmitted and scattered powers, $P_T + P_S$.

The three photodetectors were connected to separate Stanford Research Systems SR830 lock-in amplifiers. The internal oscillator of one amplifier was set to 1.243 kHz and used both to modulate the AOTF driver and to provide the reference signal to the other amplifiers, enabling simultaneous detection of the beam-monitor, reflection, and integrating-sphere signals at the modulation frequency.

### 1.1 Polarization Characterization at the Sample Plane

The polarization state delivered to the sample was characterized at the focal plane using a commercial polarization analyzer (SK010PA, Schäfter + Kirchhoff GmbH). For this measurement, the sample was replaced with the polarization-analyzer sensor, which was positioned at the focal plane of the microscope objective. The polarization state was measured independently at each excitation wavelength over the complete 500–780 nm spectral range. The orientations of the half-wave plate and quarter-wave plate used during the polarization characterization were subsequently retained for the corresponding H/V and RCP/LCP single-particle measurements, respectively.

The polarization-analyzer software directly reported the total optical intensity, $S_0$, together with the normalized Stokes parameters $S_1$, $S_2$, and $S_3$. Here, $S_1$ describes the normalized intensity imbalance between the orthogonal H- and V-polarized components, $S_2$ describes the normalized intensity imbalance between the $+45°$- and $-45°$-polarized components, and $S_3$ describes the normalized intensity imbalance between the RCP and LCP components.

Following the sign convention of the polarimeter, the normalized linear Stokes parameter was defined as

$$S_1 = \frac{I_V - I_H}{I_V + I_H}, \qquad \text{(S1)}$$

such that ideally $S_1 = +1$ for V-polarized light and $S_1 = -1$ for H-polarized light. The normalized circular Stokes parameter was defined as

$$S_3 = \frac{I_{RCP} - I_{LCP}}{I_{RCP} + I_{LCP}}, \qquad \text{(S2)}$$

such that ideally $S_3 = +1$ for RCP illumination and $S_3 = -1$ for LCP illumination, according to the handedness convention employed by the polarimeter.

As shown in Figures S1b and S1c, the measured $S_1$ and $S_3$ values remained close to their respective ideal values throughout the investigated wavelength range, demonstrating that the selected linear and circular polarization states were delivered to the sample focal plane with high fidelity.

### 1.2 Particle Localization and Polarization-Resolved Spectral Acquisition

Individual nanoparticles were located by raster-scanning the glass substrate using the three-axis piezoelectric stage at a fixed excitation wavelength close to the particle resonance. Gold nanospheres and both L- and D-helicoid particles were initially scanned using 550 nm excitation, whereas gold nanorods were scanned using 650 nm excitation. The initial raster scan produced a two-dimensional reflectance map, from which candidate nanoparticles were identified by their characteristic reflectance contrast relative to the surrounding glass substrate.

Each candidate particle was subsequently examined using a finer two-dimensional raster scan with a step size of 250 nm to obtain a higher-resolution reflectance map and accurately determine the particle position. The particle center was assigned to the position exhibiting the strongest particle-associated reflectance contrast. The excitation beam was then parked at this position while wavelength-resolved spectra were acquired. For the nanospheres and L- and D-helicoids, spectra were measured sequentially under RCP and LCP illumination. For the nanorods, spectra were measured under RCP, LCP, H, and V illumination to enable comparison of the circular- and linear-polarization responses.

For each particle, an equivalent wavelength-resolved measurement was subsequently performed on a nearby particle-free region of the same glass substrate under otherwise identical conditions. This measurement served as the local glass reference used in the reference correction described below.

### 1.3 Detector Normalization and Absorptance Calculation

Absorptance measurements were performed independently under H, V, RCP, and LCP illumination. For each polarization state p, where

$$p \in \{H, V, RCP, LCP\},$$

every detector signal was divided by the simultaneously recorded beam-monitor signal to correct for temporal variations in the laser output and wavelength-dependent variations in the AOTF transmission. The resulting beam-monitor-normalized signals are denoted $P_{R,p}$ for the reflection detector and $P_{T,p} + P_{S,p}$ for the integrating-sphere detector.

Two reference measurements were acquired for each wavelength and polarization state. The reflection reference, $P_{R,p}^{mirror}$, was measured by positioning a mirror (Thorlabs PF10-03-P01) at the sample focal plane. The mirror measurement was used as the unity reference for the reflection channel, representing the detector response when the incident light was directed entirely into the reflection channel. The integrating-sphere reference, $\left(P_{T,p}+P_{S,p}\right)^{MISS}$, was measured under the MISS condition, in which the incident beam bypassed the sample while the sample and sample holder remained inside the integrating sphere. The MISS measurement was used as the unity reference for the integrating-sphere channel, representing the detector response when the incident light entered the sphere without interacting with the sample. The mirror and MISS measurements therefore established reference responses used to calculate the reflected fraction and the combined transmitted and scattered fraction, respectively.

For the particle measurement, the reflected fraction was calculated as

$$R_p(\lambda) = \frac{P_{R,p}^{particle}(\lambda)}{P_{R,p}^{mirror}(\lambda)}. \qquad (S3)$$

The combined transmitted and scattered fraction was calculated as

$$T_p(\lambda) + S_p(\lambda) = \frac{\left(P_{T,p}+P_{S,p}\right)^{particle}(\lambda)}{\left(P_{T,p}+P_{S,p}\right)^{MISS}(\lambda)}. \qquad (S4)$$

The raw absorptance of the particle-containing region was then obtained from the optical energy balance:

$$A_{raw,p}^{particle}(\lambda) = 1 - R_p(\lambda) - T_p(\lambda) - S_p(\lambda). \qquad (S5)$$

Equivalently,

$$A_{raw,p}^{particle}(\lambda) = 1 - \frac{P_{R,p}^{particle}(\lambda)}{P_{R,p}^{mirror}(\lambda)} - \frac{\left(P_{T,p}+P_{S,p}\right)^{particle}(\lambda)}{\left(P_{T,p}+P_{S,p}\right)^{MISS}(\lambda)}. \qquad (S6)$$

For each particle, an identical reference measurement was performed on a nearby particle-free region of the same glass substrate under the same experimental conditions. The raw reference spectrum, $A_{raw,p}^{glass}$, was calculated using the same mirror and MISS normalization procedure. The final single-particle absorptance was obtained by applying the local glass reference correction:

$$A_p(\lambda) = A_{raw,p}^{particle}(\lambda) - A_{raw,p}^{glass}(\lambda). \qquad (S7)$$

This local reference correction compensates for residual wavelength- and polarization-dependent background signals and systematic offsets present in the measurement. Because the particle and glass-reference measurements were performed on nearby regions of the same substrate under otherwise identical conditions, the correction isolates the optical response associated with the individual particle. The resulting spectra are denoted $A_H$, $A_V$, $A_{RCP}$, and $A_{LCP}$.

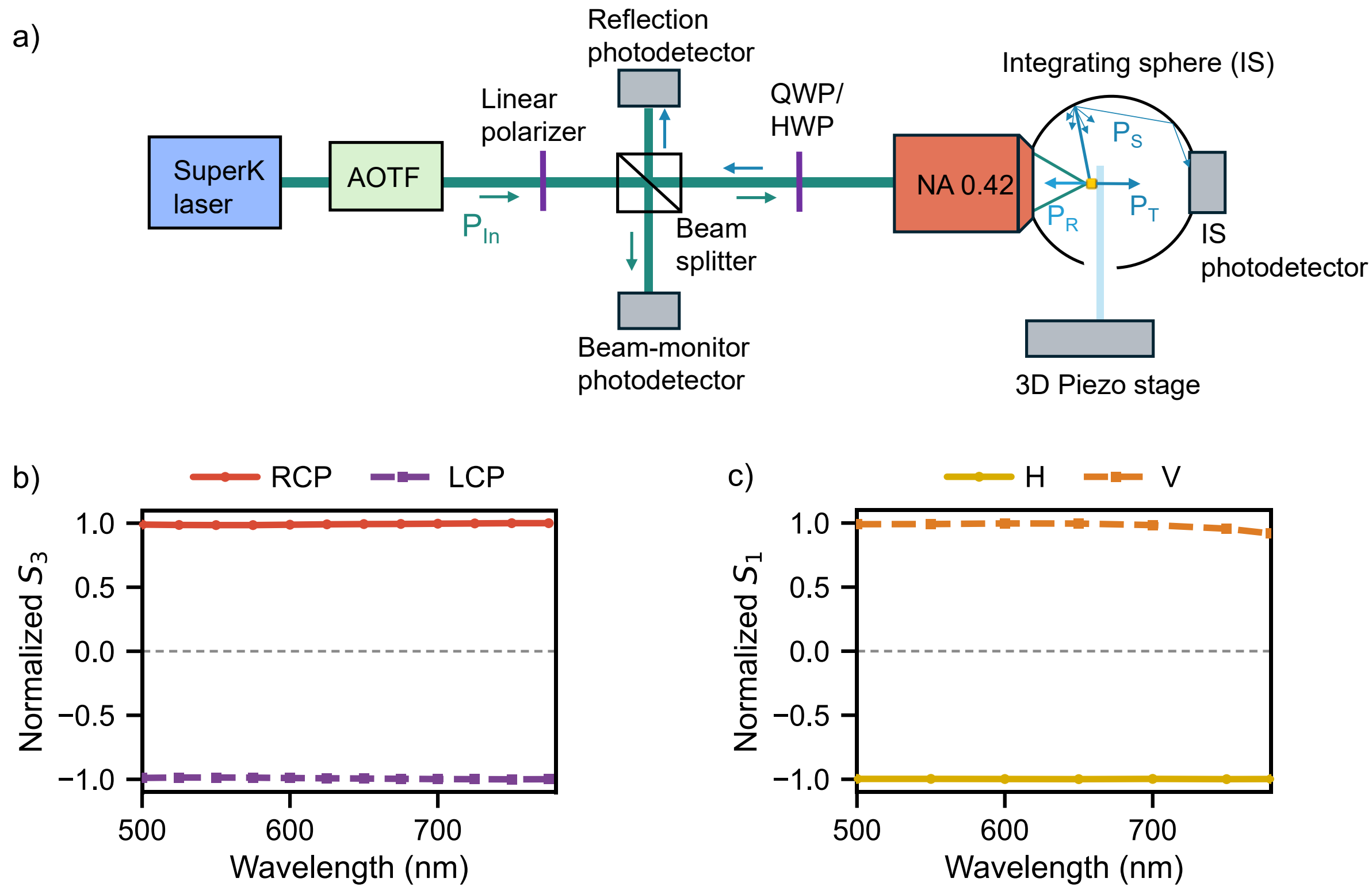

**Figure S1. Wavelength-tunable integrating-sphere microscope and polarization characterization.** (a) Schematic of the optical setup. Light from a supercontinuum laser is spectrally selected using an acousto-optic tunable filter (AOTF), linearly polarized, and converted to H, V, RCP, or LCP illumination using a half-wave plate (HWP) or quarter-wave plate (QWP). A beam-monitor photodetector records the incident-power fluctuations, while light reflected from the sample is collected through the microscope objective and directed to the reflection photodetector. Transmitted and scattered light are collected by the integrating sphere and recorded by the integrating-sphere photodetector. The sample is positioned inside the sphere using a three-axis piezoelectric stage. $P_{in}$, $P_R$, $P_T$, and $P_S$ denote the incident, reflected, transmitted, and scattered powers, respectively. (b) Wavelength-dependent normalized circular Stokes component, $S_3$, measured at the focal plane for nominal RCP and LCP illumination. (c) Wavelength-dependent normalized linear Stokes component, $S_1$, measured for nominal H and V illumination. The measured Stokes components remain close to their ideal values of $+1$and $-1$ throughout the 500–780 nm wavelength range.

## 2. Preparation of Achiral Nanorod and Nanosphere Samples

### 2.1 Preparation of Glass Substrates with Numbered Gold Grids

Glass substrates with numbered gold grids (100 µm x 100 µm grid cell) were prepared by photolithography, followed by metal deposition and lift-off. The substrates were first cleaned using a base-piranha solution and subsequently treated with oxygen plasma using a Plasmalab 80+ system (Oxford Instruments). The oxygen-plasma treatment was performed for 10 s at 50 W.

An HMDS adhesion-promoting layer was spin-coated onto the substrates using a CEE Apogee 200 Spin coater. The substrates were accelerated from 1000 rpm $s^{-1}$ to 4000 rpm, then maintained for 40 s. The HMDS-coated substrates were then baked on a hot plate at 150 °C for 1 min.

A layer of ma-N 1420 negative photoresist (Micro Resist Technology GmbH) was subsequently spin-coated by accelerating the substrates at 1000 rpm $s^{-1}$ to 3000 rpm and maintaining this speed for 45 s. The resist-coated substrates were soft-baked on a hot plate at 120 °C for 2 min.

The resist was exposed through a photolithography mask containing the numbered grid pattern using a MABA6 UV mask aligner (Suss MABA6 UV Mask aligner). Ultraviolet exposure was performed for 18 s at an intensity of 20 mW $cm^{-2}$. The exposed pattern was developed in ma-D 533S developer (Micro Resist Technology GmbH) for 90 seconds, then rinsed with deionized water.

A 5 nm chromium adhesion layer and a 50 nm gold layer were then deposited by electron-beam evaporation at a rate of 0.05 nm $s^{-1}$ using a Polyteknik E-Flex physical vapor deposition system. Finally, lift-off was performed in acetone to remove the unwanted metal. The substrates were subsequently rinsed with isopropyl alcohol, leaving the numbered gold grids patterned on the glass surface.

### 2.2 Deposition of nanospheres and nanorods on the glass substrate

Glass substrates with numbered grids were used for both nanorod and nanosphere samples. They were treated in a UV–ozone cleaner (BioForce UV/Ozone ProCleaner) for 10 min to make the surface hydrophilic. Citrate-capped gold nanorods with a nominal diameter of 40 nm and a longitudinal surface plasmon resonance centered at 650 nm were purchased from Nanopartz (A12-40-650-CIT-DIH-1-25). The nanorod stock suspension was diluted in isopropyl alcohol by mixing 1 µL stock solution with 29 µL IPA. PEG-carboxyl-functionalized 100 nm gold nanospheres were purchased from nanoComposix (San Diego, USA). Before dilution, the nanosphere stock suspension was sonicated for approximately 2 min to ensure uniform particle dispersion and was subsequently diluted by mixing 3 µL stock solution with 97 µL of IPA. For both samples, 2 µL of the diluted particle suspension was drop-cast onto the UV–ozone-treated glass substrate and allowed to dry completely under ambient conditions.

Following completion of the single-particle optical measurements, the samples were coated with a 5 nm chromium layer to improve surface conductivity for SEM imaging. The coating was applied using a Leica EM ACE600 double sputter coater under argon at a working pressure of $8 \times 10^{-3}$ mbar. Chromium was deposited at 0.4 nm $s^{-1}$ using a sputtering current of 150 mA for 12.5 s. The sample was rotated during deposition, with a target-to-sample distance of 1 mm, to promote uniform coating across the substrate.

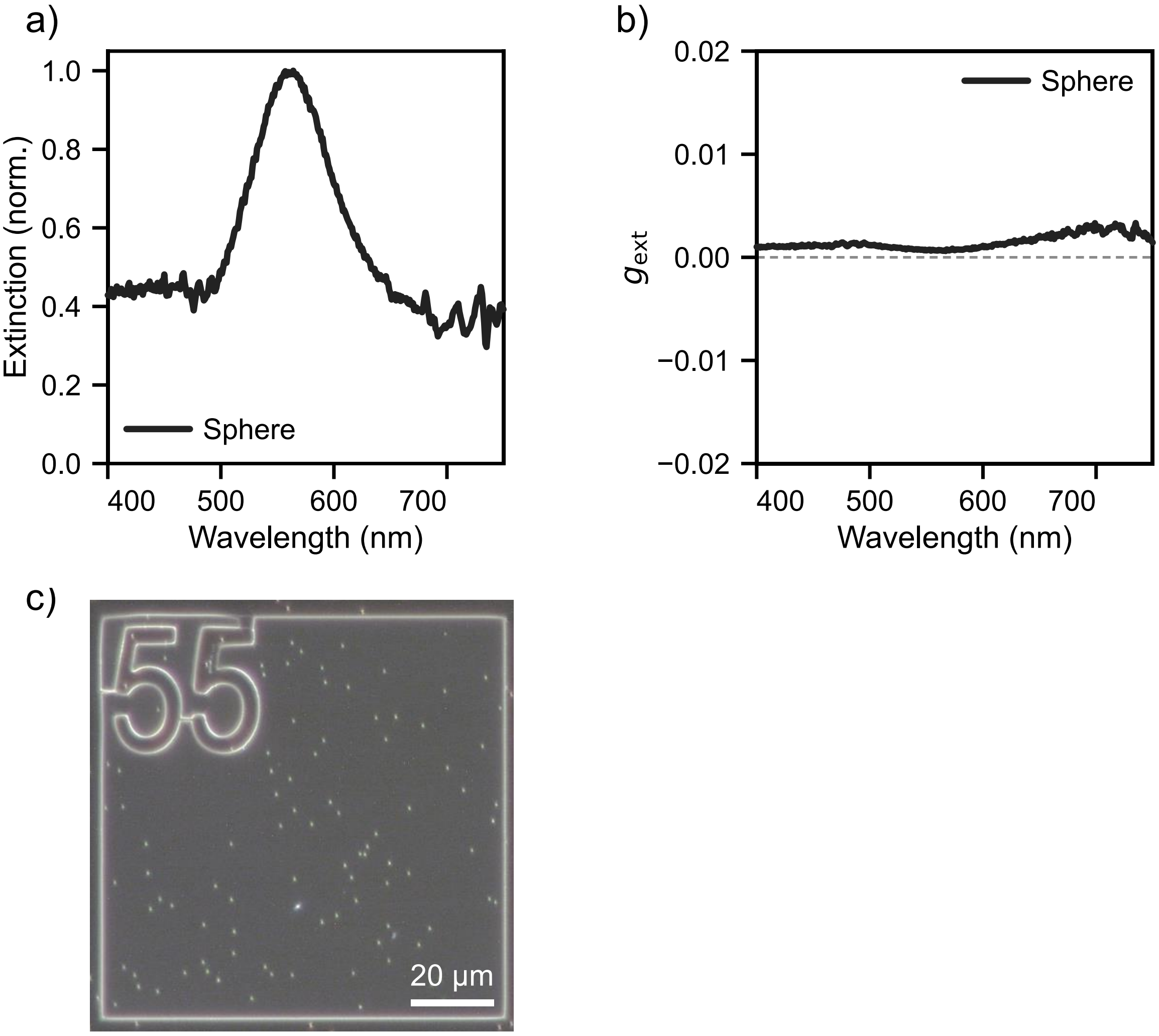

**Figure S2. Optical characterization of the achiral gold nanosphere control sample.** (a) Normalized ensemble extinction spectrum of the PEG-carboxyl-functionalized 100 nm gold nanospheres. (b) Corresponding extinction dissymmetry spectrum, $g_{ext}$, which remains close to zero across the measured wavelength range, consistent with the achiral nature of the nanospheres. (c) Representative dark-field microscopy image of the nanospheres drop-cast on the glass substrate.

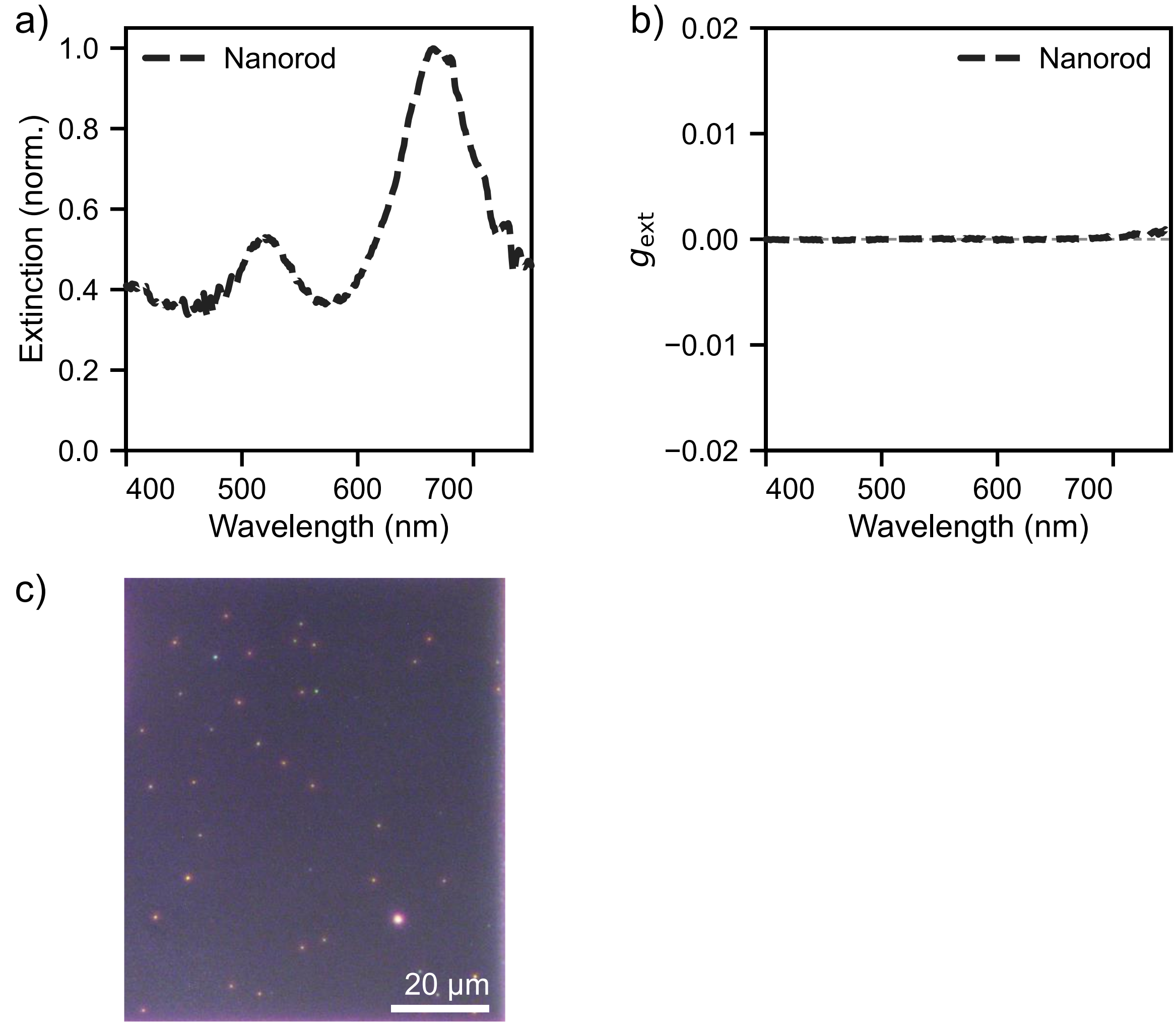

**Figure S3. Optical characterization of the achiral gold nanorod control sample.** (a) Normalized ensemble extinction spectrum of the citrate-capped gold nanorods, showing the transverse and longitudinal plasmon resonances. (b) Corresponding extinction dissymmetry spectrum, $g_{ext}$, which remains close to zero across the measured wavelength range, consistent with the achiral nature of the nanorods. (c) Representative dark-field microscopy image of the nanorods drop-cast on the glass substrate.

## 3. Population-Averaged Single Particle Achiral Responses and Additional Nanorod Control

To complement the population analysis presented in the main text, Figure S4 shows the population-averaged $A_{RCP}$ and $A_{LCP}$ spectra of the 16 gold nanospheres and 11 gold nanorods used as achiral controls. For both particle populations, the average spectra closely overlap across their respective plasmon resonances, consistent with the near-zero population-averaged absorption dissymmetry factors reported in the main text: $g_{abs} = -0.02 \pm 0.05$ for the nanospheres and $g_{abs} = 0.02 \pm 0.05$ for the nanorods. These values represent the mean and standard deviation of the peak-averaged $g_{abs}$ values obtained from the individual particles.

The gold nanorods were deposited by drop-casting and therefore adopted random in-plane orientations on the glass substrate. Depending on the orientation of an individual nanorod relative to the incident electric field, horizontally and vertically polarized light couple differently to its longitudinal plasmon mode,[2–5] producing a nonzero linear dichroism response. The nanorods were therefore used as achiral controls to evaluate whether strong linear dichroism could leak into the measured circular dichroism channel.

In the main text, horizontally and vertically oriented nanorods are shown as limiting cases in which one linear polarization couples much more strongly to the longitudinal plasmon resonance than the orthogonal polarization. Here, we additionally show a nanorod oriented diagonally relative to the horizontal and vertical polarization axes. For this intermediate orientation, both electric-field components have a nonzero projection along the nanorod long axis. Consequently, both $A_H$ and $A_V$ exhibit a pronounced longitudinal plasmon resonance, although with different amplitudes, resulting in $g_{LD} = 0.51 \pm 0.05$. Despite this pronounced linear response, the corresponding $A_{RCP}$ and $A_{LCP}$ spectra closely overlap, yielding $g_{abs} = -0.01 \pm 0.02$.

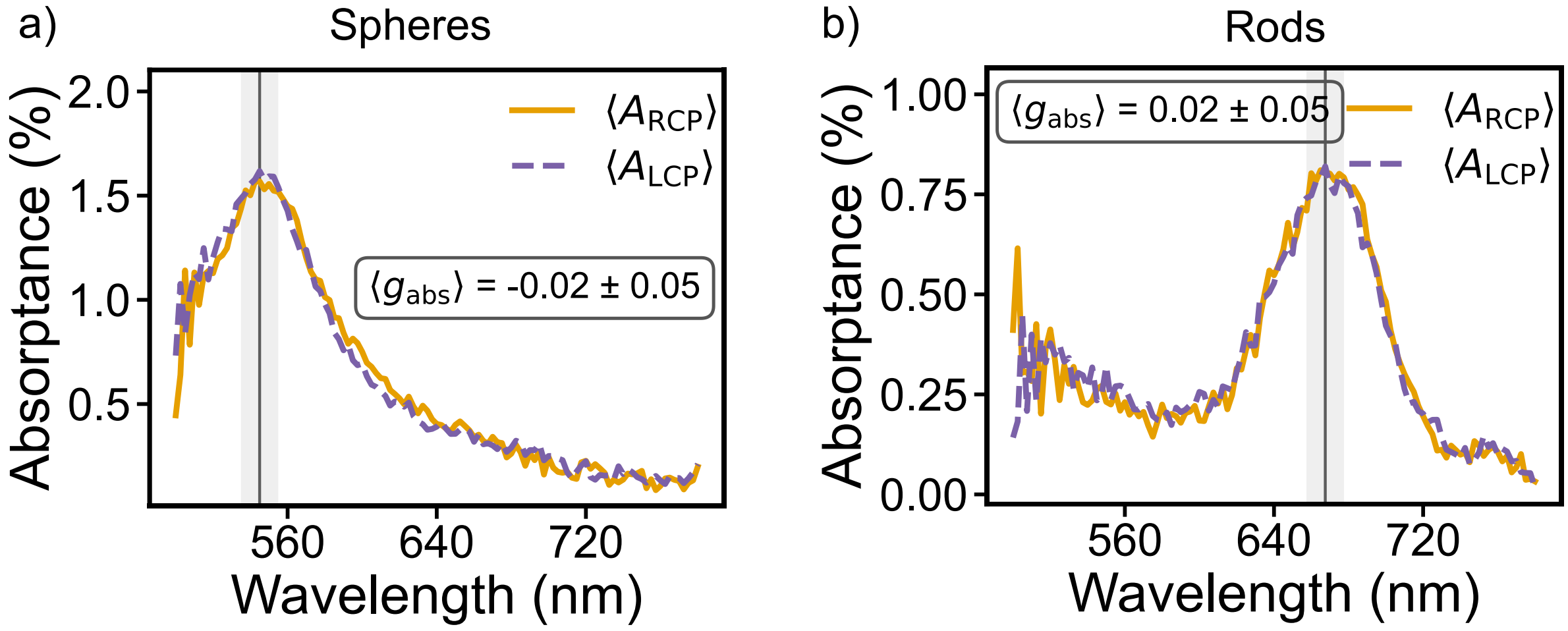

**Figure S4. Population-averaged absorptance spectra of the achiral nanoparticle controls.** (a) Average $A_{RCP}$ and $A_{LCP}$ spectra of 16 gold nanospheres. (b) Average $A_{RCP}$ and $A_{LCP}$ spectra of 11 gold nanorods. The close overlap between the two circular-polarization spectra is consistent with the near-zero population-averaged $g_{abs}$ values reported in the main text. Shaded regions indicate the 20 nm-wide spectral windows used for the peak-averaged analysis.

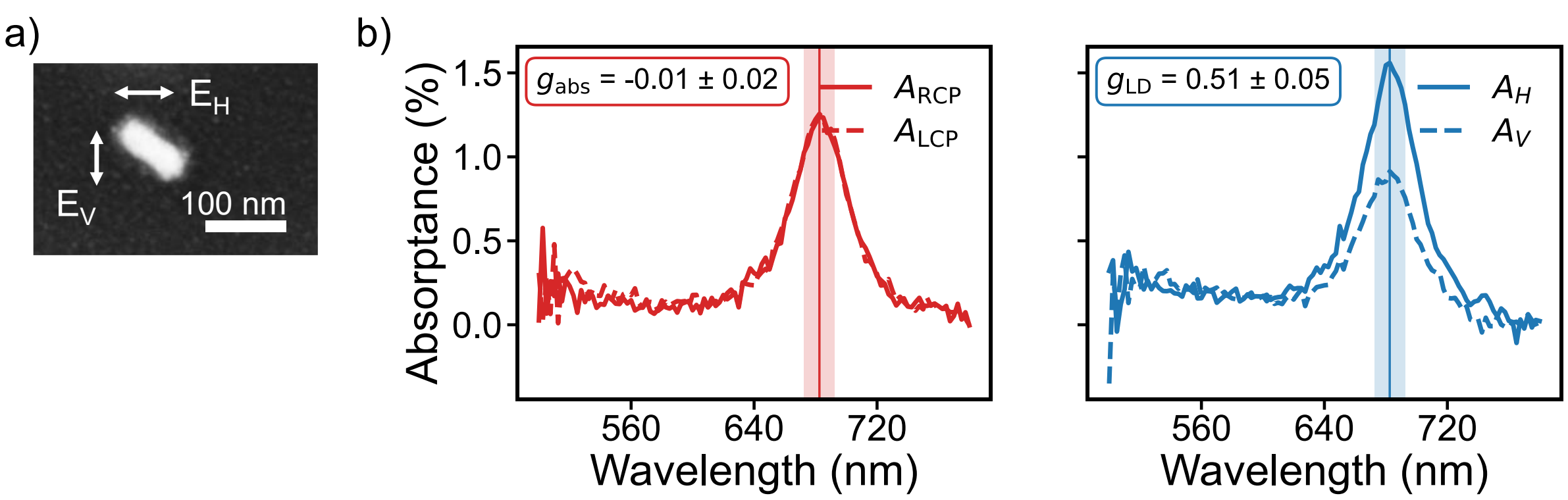

**Figure S5. Circular- and linear-polarization responses of a diagonally oriented gold nanorod.** (a) Scanning electron micrograph of a gold nanorod oriented diagonally relative to the horizontal, $E_H$, and vertical, $E_V$, polarization axes. (b) $A_{RCP}$ and $A_{LCP}$ spectra, yielding a peak-averaged $g_{abs} = -0.01 \pm 0.02$. (c) $A_H$ and $A_V$ spectra. Both linear-polarization components couple to the longitudinal resonance because of the intermediate rod orientation, resulting in $g_{LD} = 0.51 \pm 0.05$. Shaded regions indicate the 20 nm-wide spectral windows used for averaging.

## 4. Extraction of Control-Particle $g_{abs}$ and $g_{LD}$

### 4.1 Peak-Averaged Dissymmetry Factors

For each nanorod, the circular- and linear-polarization resonance wavelengths were determined independently from the maxima of the corresponding polarization-averaged absorptance spectra,

$$A_{circ} = \frac{A_{RCP} + A_{LCP}}{2}$$

and

$$A_{lin} = \frac{A_{H} + A_{V}}{2}.$$

The representative $g_{abs}$ and $g_{LD}$ values were obtained by averaging the corresponding dissymmetry-factor spectra within a 20 nm-wide window centered on their independently determined resonance wavelengths.

For each nanosphere, the resonance wavelength was identified from the maximum of $A_{circ}$. The representative $g_{abs}$ value was then obtained by averaging within a 20 nm-wide window centered on this wavelength.

For all control particles, the standard deviation within the averaging window was used to describe the spectral variation around the resonance. No spectral smoothing or line-shape fitting was applied.

### 4.2 Achiral-Control Threshold

The peak-averaged $g_{abs}$ values obtained from all measured nanorods and nanospheres were combined into a single achiral-control population. The mean, $\langle g_{abs} \rangle_{control}$, and standard deviation, $\sigma_{control}$, of the pooled distribution were calculated, and the achiral-control interval was defined from the pooled mean ± 1 SD. The resulting reference range used throughout the analysis was

$$-0.06 \le g_{abs} \le +0.06$$

Values within this interval were classified as within control. Pooling the nanorod and nanosphere measurements provided a control-derived experimental baseline that accounts for the observed variation across achiral particles with different morphologies.

## 5. Gold Helicoid Synthesis and Sample Preparation

### 5.1 Helicoid III Synthesis

#### Materials and Chemicals

All chemicals were purchased from the suppliers specified below and used without further purification. Cetyltrimethylammonium chloride (CTAC; 25 wt% aqueous solution; Merck, Cat. No. 292737-500ML, CAS No. 112-02-7), cetyltrimethylammonium bromide (CTAB; molecular biology grade; Merck, Cat. No. H6269-100G, CAS No. 57-09-0), gold(III) chloride trihydrate ($HAuCl_4$; ≥99.9% metals basis; Merck, Cat. No. 520918-1G, CAS No. 16961-25-4), sodium borohydride ($NaBH_4$; Merck, CAS No. 16940-66-2), potassium iodide (KI; ≥99.99% metals basis; Merck, Cat. No. 204102-10G, CAS No. 7681-11-0), L-(+)-ascorbic acid (Merck, Cat. No. 1.00468.0100, CAS No. 50-81-7), reduced L-glutathione (L-GSH; Merck, CAS No. 70-18-8), and reduced D-glutathione (D-GSH; Angene Chemicals) were used in the synthesis.

All solutions were prepared using Milli-Q water and used within 1 h unless otherwise stated.

**Synthesis Procedure**

Helicoid III nanoparticles, hereafter abbreviated as H3, were synthesized following previously reported procedures.[6,7] The synthesis consisted of three stages. First, small gold nanospheres were synthesized as crystallization seeds and are hereafter referred to as AuNS seeds. The AuNS seeds were subsequently grown into gold nanooctahedra, referred to as AuNO. In the final stage, the AuNO seeds were overgrown in the presence of either L-GSH or D-GSH as the chiral growth-directing agent, producing the corresponding L-H3 and D-H3 nanoparticles.

**Step 1. AuNS Seed Synthesis**

An AuNS growth solution was prepared in a 20 mL glass vial by sequentially adding a magnetic stir bar, 8.43 mL of Milli-Q water, 1.32 mL of 25 wt% aqueous CTAC, and 0.25 mL of 10 mM $HAuCl_4$. Subsequently, 0.45 mL of a freshly prepared 20 mM $NaBH_4$ solution in ice-cold Milli-Q water was rapidly added. The $NaBH_4$ solution was used immediately after preparation. The reaction mixture was stirred for 10 s, after which stirring was stopped, and the magnetic stir bar was removed. The reaction was allowed to proceed for 1 h at a temperature below 25 °C. The resulting AuNS seed solution exhibited a dark-brown color.

**Step 2. Small-Scale AuNO Synthesis**

Two growth solutions, denoted A and B, were prepared separately in 20 mL glass vials. Each solution contained 8.15 mL of Milli-Q water, 1.32 mL of 25 wt% aqueous CTAC, 0.25 mL of 10 mM $HAuCl_4$, and 5 µL of 10 mM KI in Milli-Q water. Solution B was incubated at 30 °C.

The subsequent steps were performed in rapid sequence while strictly adhering to the indicated timings. A 220 µL aliquot of 40 mM aqueous ascorbic acid was added to each of solutions A and B, followed by stirring for 10 s. Subsequently, 40 µL of the AuNS seed solution was injected into solution A, which was stirred for 7 s. During this period, the solution changed from colorless to fuchsia. An aliquot of 900 µL of solution A was then injected into solution B no later than 15 s after the addition of the AuNS seed solution to solution A. Solution B was allowed to react for 15 min at 30 °C.

The resulting AuNO particles were isolated by centrifugation at 3515 g for 15 min and redispersed in 50 mL of 3 mM aqueous CTAC. The resulting AuNO suspension exhibited a deep-purple color. The wavelength of maximum absorbance of the AuNO suspension was required to lie between 573 and 582 nm for subsequent H3 synthesis.

**Step 3. L-H3 and D-H3 Synthesis**

For the synthesis of the L-H3 and D-H3 nanoparticles, a growth solution was prepared in a 10 mL glass vial maintained at 30 °C. The growth solution contained 3.95 mL of Milli-Q water, 0.80 mL of 100 mM CTAB, and 100 µL of 10 mM $HAuCl_4$. Under mild magnetic stirring, 475 µL of 100 mM aqueous ascorbic acid was added, followed rapidly by 5 µL of 5.5 mM L-GSH or D-GSH and 50 µL of the AuNO suspension. Before use, the concentration of the AuNO suspension was adjusted such that its maximum absorbance, $A_{max}$, was 0.315.

The reaction was allowed to proceed for 2 h at 30 °C. The resulting H3 nanoparticles were recovered by centrifugation at 2000 g for 3 min. The supernatant was discarded, and the particle pellet was redispersed in 0.5 mL of 1 mM aqueous CTAB.

### 5.2 Bulk-Solution Optical and Chiroptical Characterization

Bulk optical and chiroptical measurements were performed using a Jasco J-1500 CD spectrophotometer and 10 mm path-length quartz cuvettes (QS high-precision cells, Hellma Analytics). Milli-Q water was used as the baseline for all measurements. The measurement parameters were as follows: bandwidth, 4 nm; wavelength range, 300–800 nm; CD sensitivity, 2000 mdeg/1.0 dOD; scanning speed, 100 nm $min^{-1}$; data integration time, 0.5 s; data interval, 1 nm; number of accumulations, 5; and temperature, 25 °C.

The bulk extinction and CD spectra of the achiral-control gold nanorods and gold nanospheres are shown in Figure S2 and Figure S3. The H3 samples were characterized spectroscopically after each stage of the synthesis. Extinction spectra were recorded for the AuNS seed solution and AuNO suspension, whereas both extinction and CD spectra were recorded for the final L-H3 and D-H3 products. Representative spectra of the AuNS and AuNO intermediates have been reported previously.[6,7] The bulk extinction and CD spectra of the L-H3 and D-H3 batches used in this study are shown in Figure 3a and Figure 3b of the Main text.

### 5.3 Preparation of Helicoid Samples for Single-Particle Measurements

The L- and D-helicoid samples were deposited on glass substrates containing numbered gold grids prepared as described in Section 2.1. Before deposition, the substrates were treated in a UV–ozone cleaner (BioForce UV/Ozone ProCleaner) for 10 min to render the glass surface hydrophilic. The L- and D-helicoid suspensions were diluted separately in isopropyl alcohol by mixing 3 µL of the corresponding helicoid suspension with 47 µL of IPA. The diluted suspensions were then drop-cast onto separate substrates and allowed to dry completely under ambient conditions. The deposited samples contained spatially isolated particles with interparticle separations greater than 3 µm, enabling individual particles to be measured without significant contributions from neighboring particles.

Following completion of the single-particle optical measurements, the samples were coated with a 10 nm chromium layer to improve surface conductivity for SEM imaging. The coating was applied using a Leica EM ACE600 double sputter coater under argon at a working pressure of $8 \times 10^{-3}$ mbar. Chromium was deposited at 0.4 nm $s^{-1}$ using a sputtering current of 150 mA for 25 s. The sample was rotated during deposition, with a target-to-sample distance of 1 mm, to promote uniform coating across the substrate.

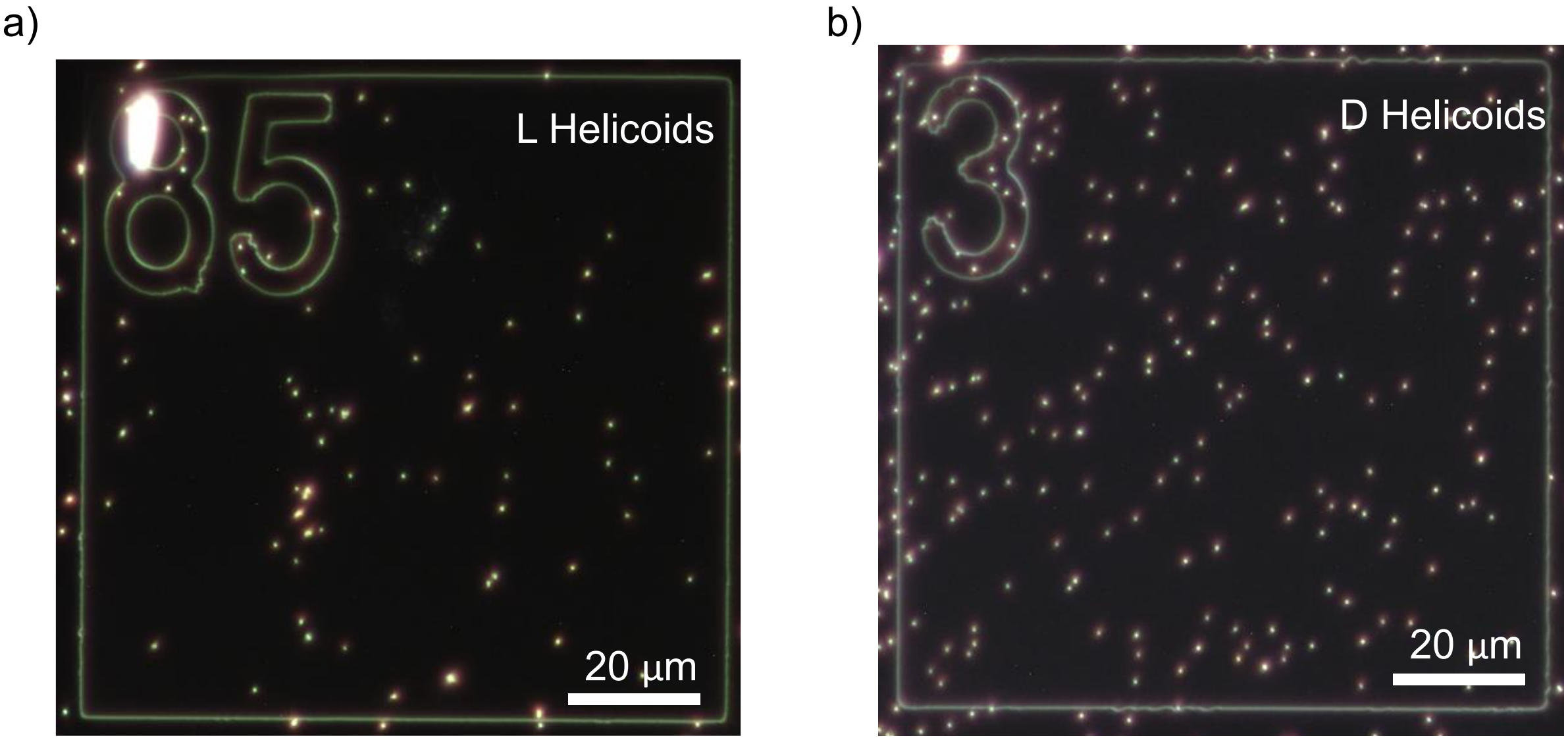

**Figure S6. Dark-field microscopy images of the helicoid samples prepared for single-particle measurements.** Representative dark-field microscopy images of the deposited (a) L-helicoid and (b) D-helicoid samples on glass substrates containing numbered gold grids. The deposition conditions produced spatially isolated particles with interparticle separations greater than 3 µm, suitable for single-particle optical measurements and subsequent particle relocation.

## 6. Extraction of Particle-Level $g_{abs}$ and Population Analysis

For each helicoid, the wavelength-dependent absorption dissymmetry factor was calculated as

$$g_{abs}(\lambda) = \frac{2[A_{RCP}(\lambda) - A_{LCP}(\lambda)]}{A_{RCP}(\lambda) + A_{LCP}(\lambda)}.$$

According to this convention, positive $g_{abs}$ indicates stronger absorption under RCP illumination, whereas negative $g_{abs}$ indicates stronger absorption under LCP illumination. No spectral smoothing was applied before analysis.

For each of the L- and D-helicoid populations, a spectral search range was predefined to encompass red-shifted chiroptically active features associated with the characteristic opposite-sign population-average responses around 660 nm. The same search range was then applied consistently to all particles within the corresponding population. Within this range, positive local maxima and negative local minima in the unsmoothed $g_{abs}(\lambda)$ spectrum were identified as candidate response features. Candidate features were required to have a minimum prominence of 0.03 and a minimum spectral width of 5 nm. Spectral data extending 25 nm beyond the search-range boundaries were retained only to allow reliable determination of the prominence and width of features close to the boundaries; candidate centers were required to lie within the predefined search range.

For each candidate feature, the mean $g_{abs}$ was calculated over a 20 nm-wide spectral window centered on the candidate feature. Candidate features were ranked using

$$\text{selection score} = \text{prominence} \times |\overline{g_{abs}}|,$$

where $\overline{g_{abs}}$ is the mean $g_{abs}$ within the corresponding 20 nm-wide spectral window. The candidate with the highest selection score was used to define the characteristic particle-level response. The reported $g_{abs}$ value corresponds to the mean within this spectral window, and the associated uncertainty is the standard deviation of the $g_{abs}$ values within the same window.

Representative examples of the particle-level $g_{abs}$ selection and corresponding averaging windows are shown for L-helicoids in Figures 5b–e of the main text and for D-helicoids in Figure S7.

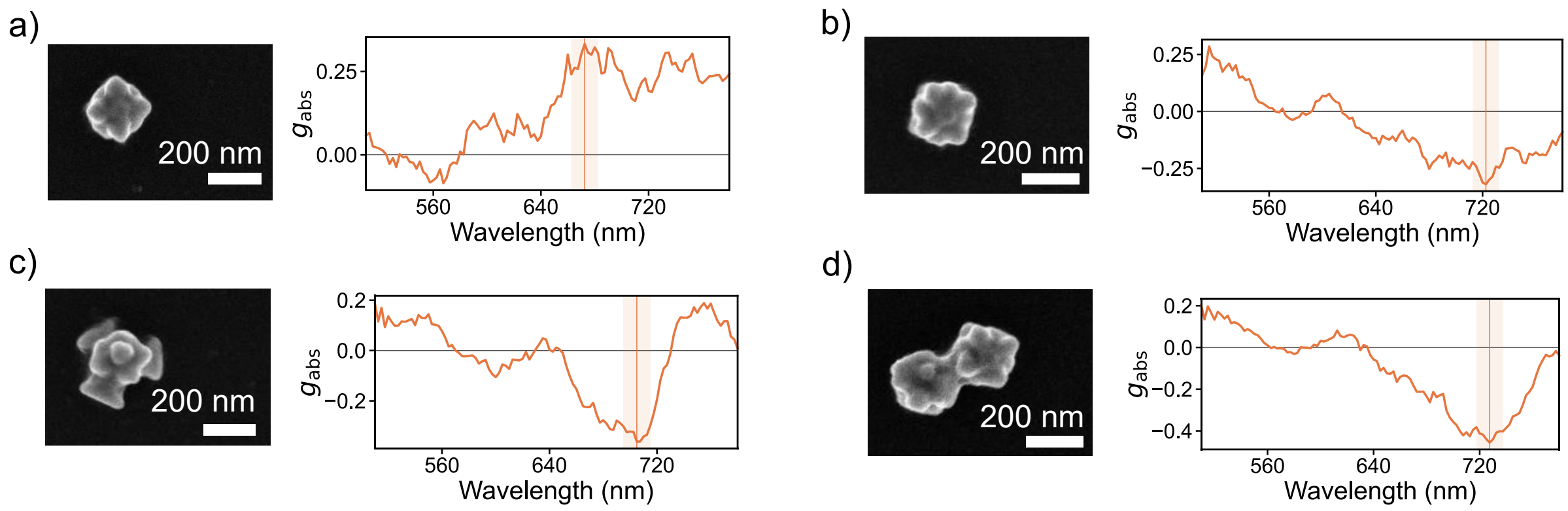

**Figure S7. Representative SEM-correlated D-helicoids and their particle-level absorption CD responses.** (a–d) SEM images and corresponding spectra of four representative D-helicoids. Shaded regions indicate the 20 nm-wide spectral windows selected by the feature-selection procedure described in Section 6 and used to determine the particle-level $g_{abs}$ values.

The population analysis included all 87 L-helicoids and 96 D-helicoids. The mean particle-level $g_{abs}$ values were −0.11 for the L-helicoids and +0.10 for the D-helicoids. The population means were compared using a two-sided Welch independent-samples t-test, revealing a statistically significant difference between the L- and D-helicoids ($p = 7.99 \times 10^{-5}$).

A common achiral-control reference range of

$$-0.06 \leq g_{abs} \leq +0.06$$

was used to classify the helicoid responses. Responses were classified as expected sign, within control, or opposite sign relative to the population-average chiroptical tendency. For L-handed helicoids, responses below −0.06 were classified as expected-sign responses, and those above +0.06 as opposite-sign responses. For D-helicoids, responses above +0.06 were classified as expected sign and responses below −0.06 as opposite sign. Responses between −0.06 and +0.06, inclusive, were classified as within control

## 7. SEM-Correlated Morphology Classification and Statistical Testing

SEM images of the optically measured L-helicoids were correlated with the corresponding single-particle optical measurements using the grid and particle identifiers. Morphology classification was performed without access to the optical spectra or $g_{abs}$ values, ensuring that the assignments were independent of the measured optical response.

Particles were classified according to their projected SEM morphology as Characteristic singles, Atypical singles, or Aggregates. Characteristic singles displayed the well-resolved four-armed, pinwheel-like morphology characteristic of an individual L-helicoid.[6,7] Atypical singles were spatially isolated particles for which this characteristic projected morphology was incomplete, distorted, or insufficiently resolved. Assemblies containing two or more particles were classified as Aggregates.

Morphology assignments were obtained for 81 of the 87 optically analyzed L-helicoids, corresponding to a correlation rate of 93.1%. The correlated population comprised 39 Characteristic singles, 28 Atypical singles, and 14 Aggregates.

Characteristic singles had a mean $g_{abs}$ of −0.11, with 59% expected-sign, 5.1% within-control, and 35.9% opposite-sign responses. Atypical singles had a mean $g_{abs}$ of −0.06, with 57.1% expected-sign, 3.6% within-control, and 39.3% opposite-sign responses. Aggregates had a mean $g_{abs}$ of −0.18, with 64.3% expected-sign, 7.1% within-control, and 28.6% opposite-sign responses.

The particle-level $g_{abs}$ distributions were compared among the three morphology classes using a Kruskal–Wallis test. No statistically significant difference was detected among Characteristic singles, Atypical singles, and Aggregates ($p = 0.549$).

The association between morphology class and the occurrence of an opposite-sign response was evaluated using a two-sided Fisher–Freeman–Halton exact test for the corresponding $3 \times 2$ contingency table. Expected-sign and within-control responses were combined as not opposite-sign for this analysis. No statistically significant association was observed between morphology class and opposite-sign behavior ($p = 0.825$).

Opposite-sign responses persisted after progressively restricting the analysis to isolated and morphologically characteristic particles. Among the 67 isolated singles, 25 particles (37.3%) exhibited an opposite-sign response. Even when the analysis was restricted to the 39 Characteristic singles, 14 particles (35.9%) remained opposite-sign. Thus, opposite-sign absorption CD responses cannot be attributed exclusively to particle aggregation or to conspicuous deviations from the characteristic projected helicoid morphology. SEM images and morphology assignments for all 81 SEM-correlated L-helicoids are provided in Section 8.

## 8. SEM Catalogue of Morphology-Classified L-Helicoids

The complete SEM catalogue of the 81 L-helicoids included in the morphology-resolved analysis is shown below. Particles are labelled with their corresponding particle identifiers and morphology assignments: Characteristic single (N=39), Atypical single (N=28), or Aggregate (N=14). Scale bar: 200 nm.

Characteristic Singles (N=39)

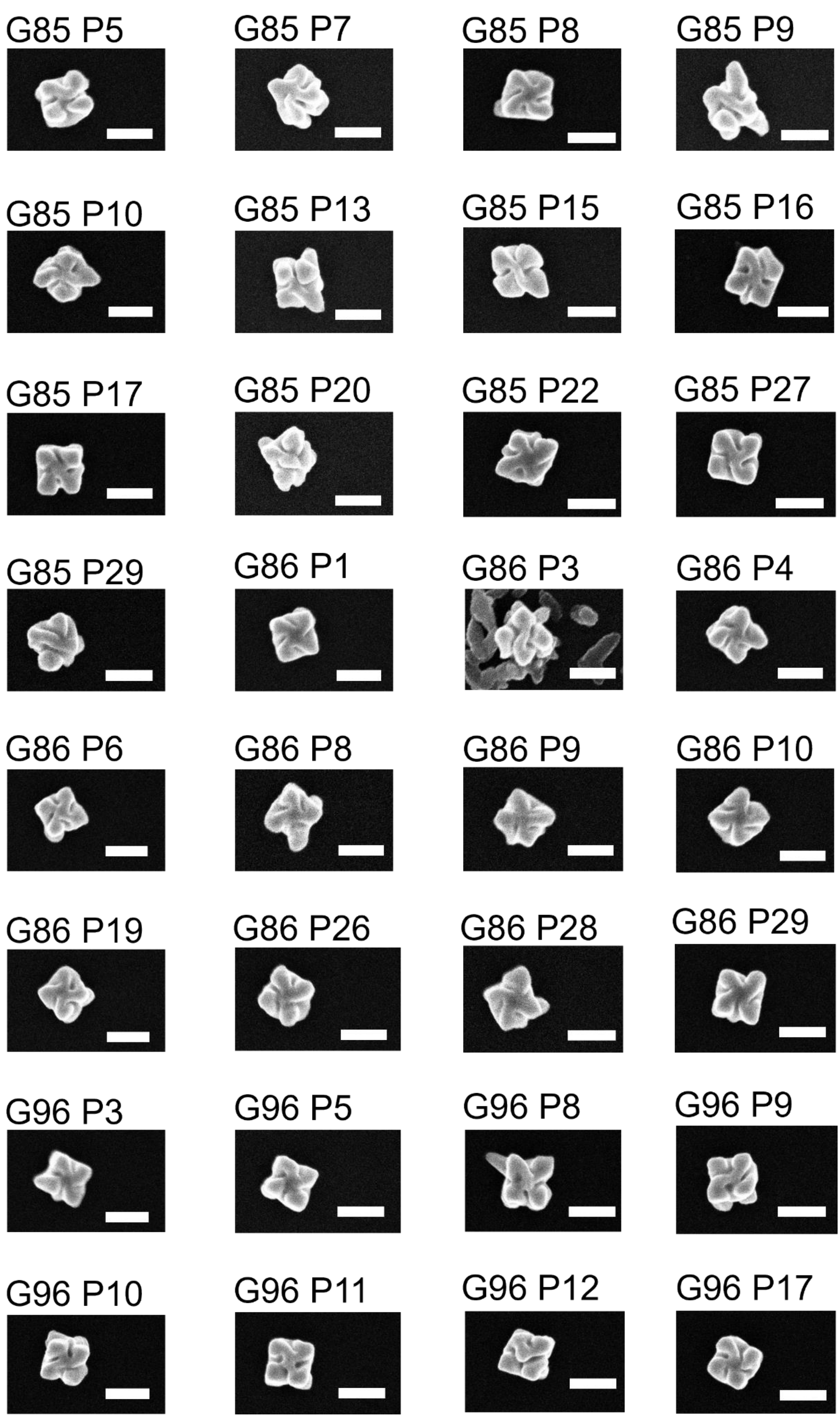

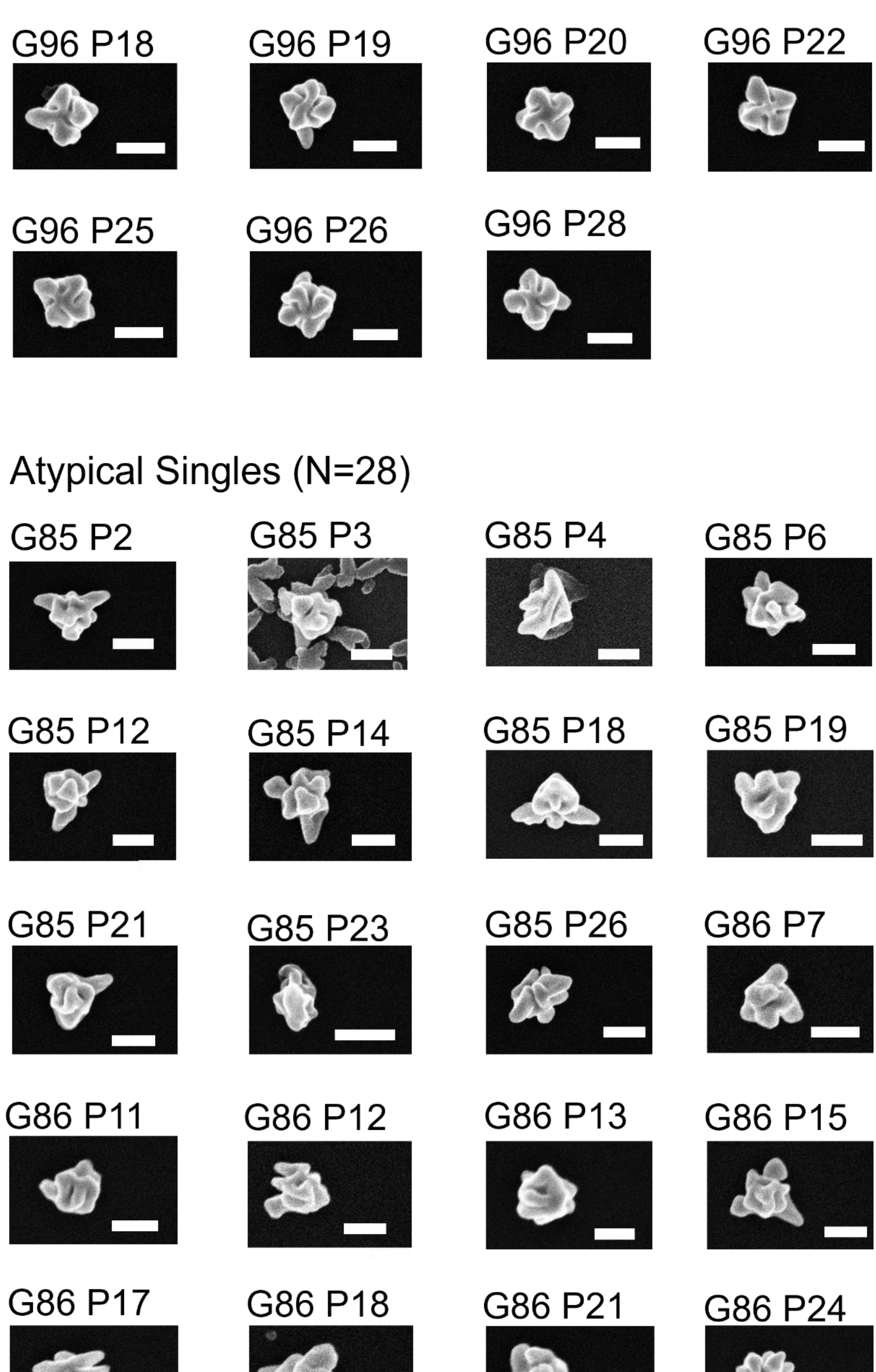
G96 P18
G96 P19
G96 P20
G96 P22
G96 P25
G96 P26
G96 P28
Atypical Singles (N=28)
G85 P2
G85 P3
G85 P4
G85 P6
G85 P12
G85 P14
G85 P18
G85 P19
G85 P21
G85 P23
G85 P26
G86 P7
G86 P11
G86 P12
G86 P13
G86 P15
G86 P17
G86 P18
G86 P21
G86 P24

G86 P27

G96 P2

G96 P7

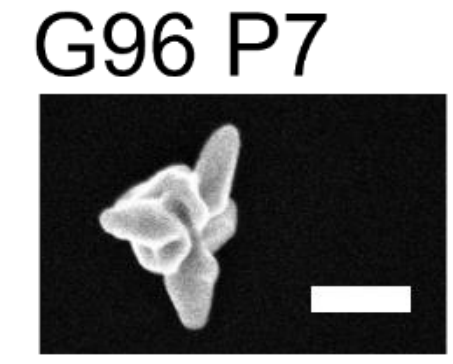

G96 P13

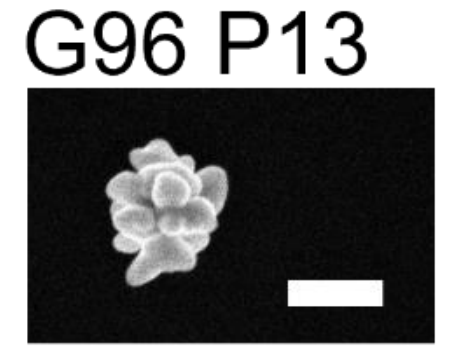

G96 P16

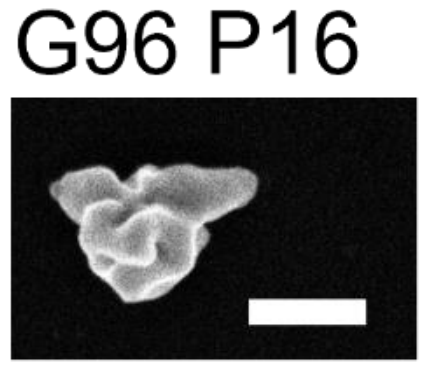

G96 P21

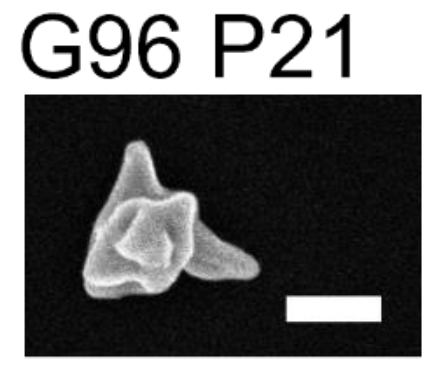

G96 P24

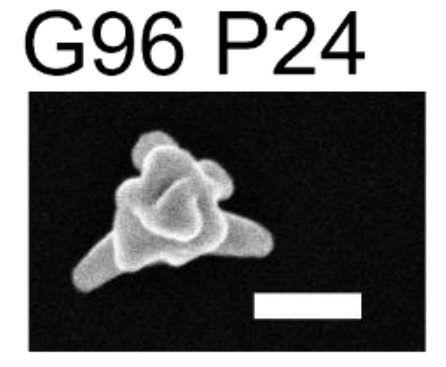

G96 P27

## Aggregates (N=14)

G85 P1

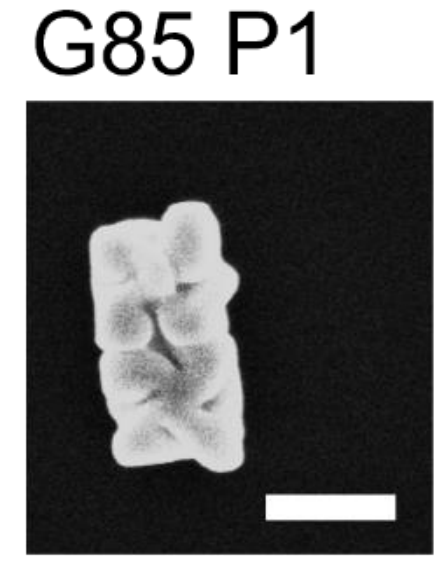

G85 P11

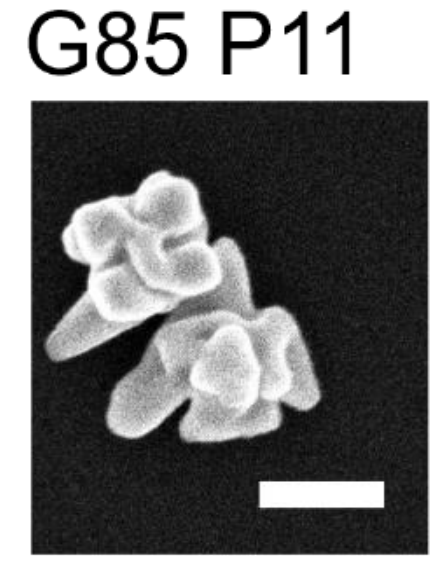

G85 P24

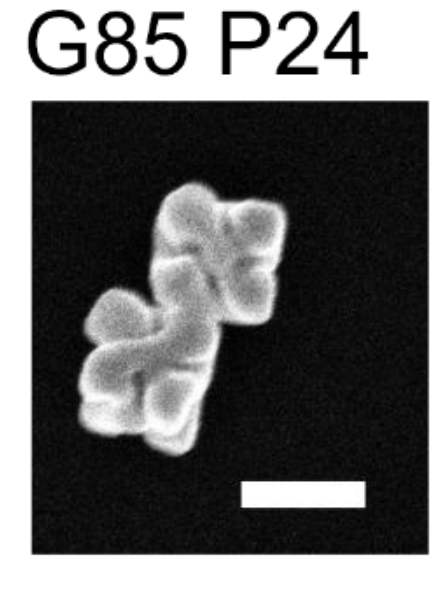

G85 P25

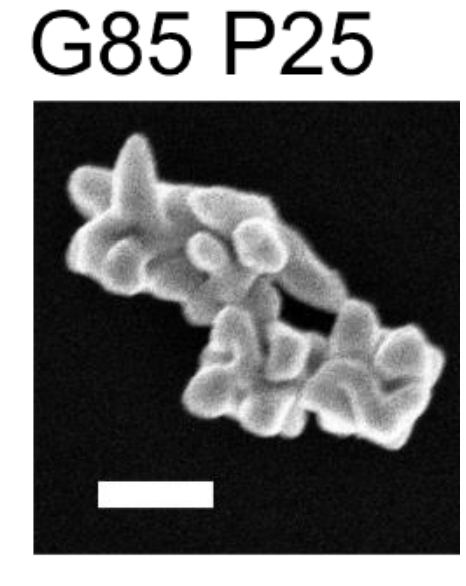

G86 P2

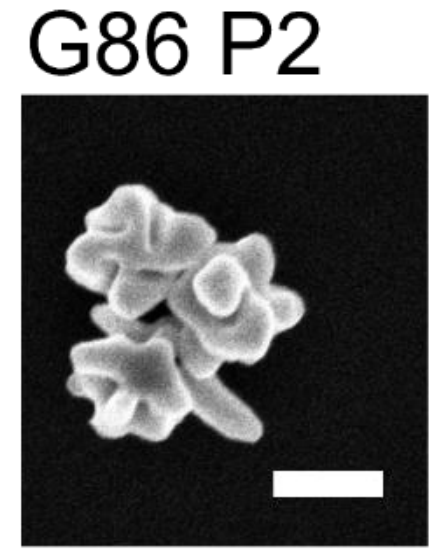

G86 P14

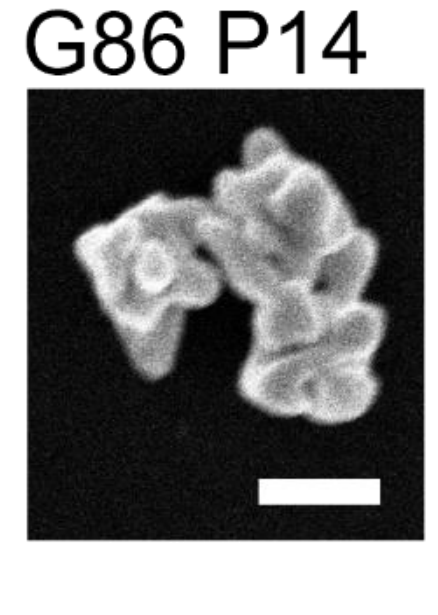

G86 P20

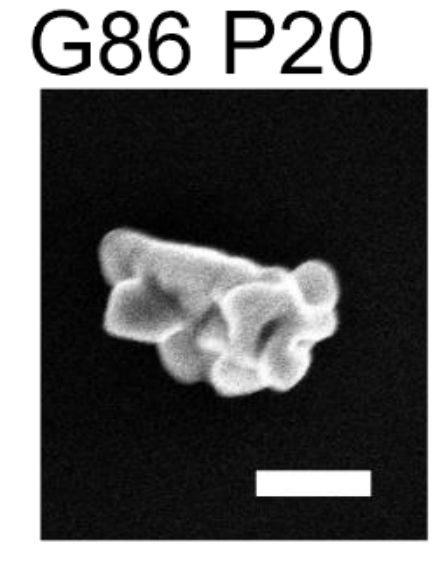

G86 P23

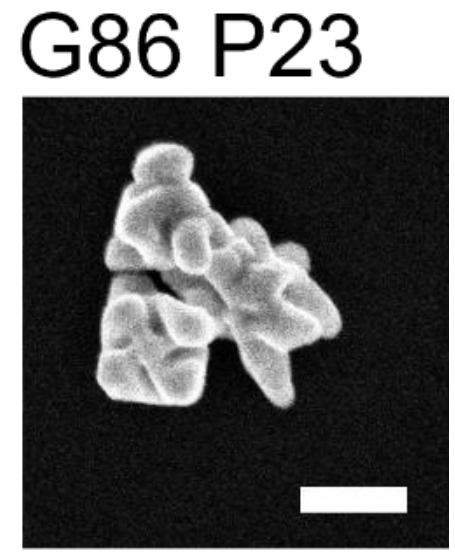

G86 P25

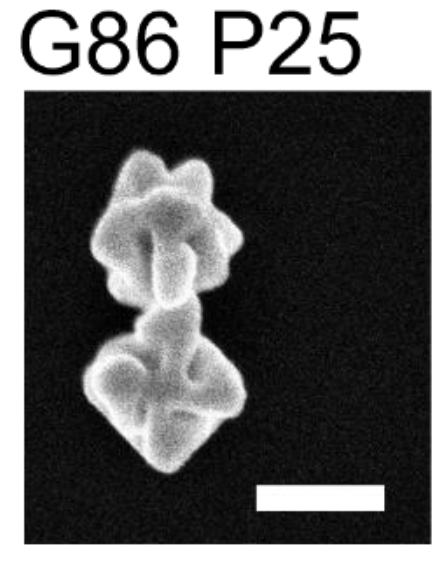

G96 P1

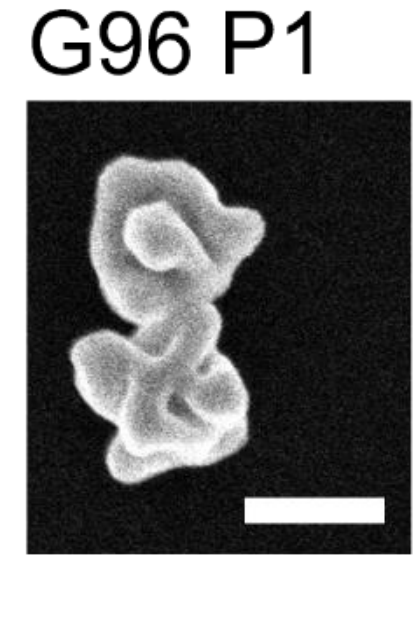

G96 P4

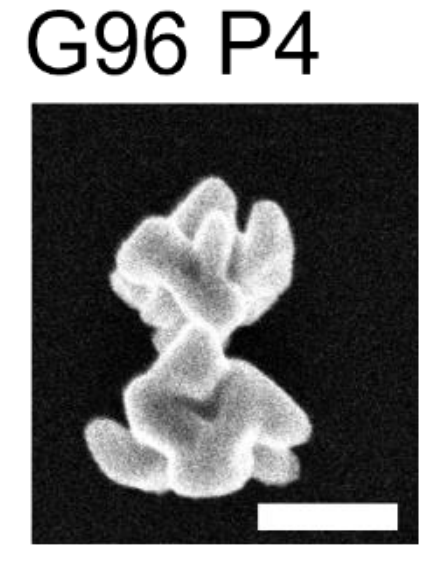

G96 P6

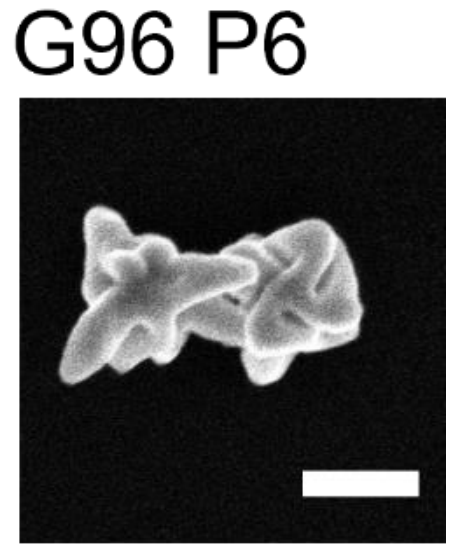

G96 P23

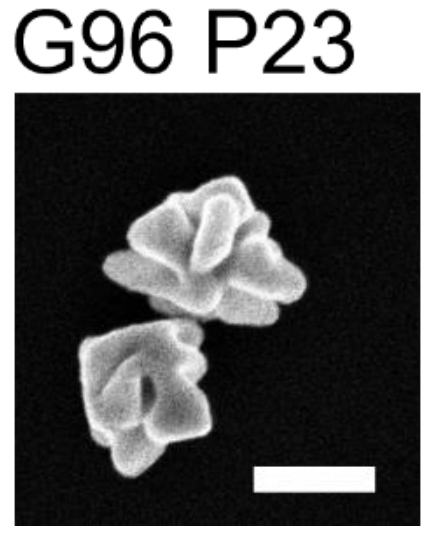